\documentclass[11pt]{article}

\usepackage{cite}
\usepackage{booktabs}
\usepackage{graphicx}
\usepackage{amsmath,amssymb}
\usepackage{array}
\usepackage{tabularx}
\usepackage{url}
\usepackage{xcolor}
\usepackage{listings}
\usepackage{microtype}
\usepackage[T1]{fontenc}
\usepackage{mathptmx}
\usepackage[letterpaper,margin=1in]{geometry}
\usepackage[hidelinks]{hyperref}

\newcolumntype{Y}{>{\raggedright\arraybackslash}X}

\begin{document}

\title{Skillware: A Software Ontology and Engineering Lifecycle for Persistent Behavioral Artifacts}

\author{
Haodi Fan\\
\texttt{anthonyfan@metainflow.cn}
\and
Zucong Lan\\
\texttt{neillan@metainflow.cn}
}
\date{}

\maketitle

\begin{abstract}
Agent Skills have become persistent behavioral artifacts across independent AI agent systems. They combine natural-language task specifications with metadata and optional references, scripts, assets, hooks, package manifests, tests, and companion interfaces. Existing studies explain how Skills are specified, executed, maintained, and evolved, but lack an ontology that defines these artifacts as independent software objects. This paper introduces \emph{Skillware} as the software abstraction that extends software engineering to persistent Behavioral Artifacts in agent systems. A Skill Artifact specifies reusable task behavior; a Skillware Unit manages that artifact as software through an independent identity and lifecycle. A compatible Agent Host activates the unit for runtime interpretation. Three necessary conditions operationalize category membership: behavioral primacy, independent software identity, and an Agent Host execution relationship. Lifecycle Continuity records whether the same unit identity persists through update, maintenance, rollback, and removal as a separate software-grade property. Evidence combines the Agent Skills specification, a frozen corpus of 138,133 content-deduplicated \texttt{SKILL.md} records associated with 20,556 repository identifiers, independent empirical studies, 15 category-boundary cases, and 13 fixed-revision engineering implementations. The evidence establishes a recurring artifact envelope, separable software identities, documented or reconstructed activation paths, and lifecycle engineering pressure. Skillware provides the software ontology and engineering lifecycle through which agent capabilities can become identifiable, composable, and maintainable software artifacts with an explicit basis for future evolution.
Public design-pattern and evidence materials are available at \url{https://github.com/MetaInFLow/skillware-patterns}.
\end{abstract}

\begin{center}
\end{center}

\section{Introduction}

Software abstractions determine how intended behavior is represented, packaged, executed, and changed. Code-centric software translates intent through formal languages into source code realized by a runtime. Model-centric intelligence uses data, objectives, architectures, and optimization to produce parameters that shape inference behavior \cite{rumelhart1986backprop,lecun2015deep,bommasani2021foundation}. Agent systems expose a third behavioral substrate. Developers can persist procedures, constraints, policies, examples, and evaluation criteria in natural language, and an agent system interprets this source within an active task. Formal code, learned parameters, and natural-language behavioral source now coexist as distinct contributors to system behavior.

Software engineering repeatedly responds to a new computational object by introducing an abstraction around it. Modules organize procedures and isolate change, components organize deployable services and interfaces, and model abstractions organize learned representations \cite{parnas1972modules,szyperski2002component,bommasani2021foundation}. Agent systems introduce persistent behavioral artifacts that require a corresponding software abstraction: the source must acquire an identity, a boundary, an execution relationship, and a way to persist through change. Skillware names that missing software ontology.

Agent Skills make this authoring surface observable as an artifact. The Agent Skills specification defines a directory with a required \texttt{SKILL.md}, YAML metadata, Markdown instructions, and optional \texttt{scripts/}, \texttt{references/}, \texttt{assets/}, and additional files \cite{agentskills2025specification}. Independent systems expose compatible Skill discovery and loading paths, including Codex, Claude Code, GitHub Copilot, Hermes, and OpenClaw \cite{openai2026skills,anthropic2026skills,github2026copilotskills,hermes2026skills}. Public projects extend the directory with hooks, agent-system-specific bindings, plugin manifests, test suites, release records, local services, and browser companions \cite{obra2026superpowers,garrytan2026gstack,affaan2026ecc,calesthio2026openmontage}. The resulting unit can be acquired, installed, activated, used, updated, maintained, and removed.

This phenomenon has outgrown the vocabulary of a prompt file. Prompt research explains model conditioning and structured interaction \cite{reynolds2021prompt,liu2023prompting,beurerkellner2023prompting}. Agent research explains reasoning, action, observation, and tool use \cite{yao2023react}. Skill research now covers architecture, acquisition, security, curation, runtime requirements, reuse, maintenance, quality, verification, and production governance \cite{xu2026agentskills,zhou2026agentskills,ouyang2026skillos,skillsnewapps2026,registrytorepository2026,anatomytosmells2026,metere2026verifiableskills,xu2026skillfab}. These studies establish many properties of the object while using different units of analysis. A shared software abstraction must connect the Skill-centered behavioral source, the independently managed unit, the application that activates it, the runtime that interprets it, and the evidence generated by execution.

We use \emph{Skillware} for that category. Our contribution concerns a cross-project operational definition and evidence protocol. We make no claim to word coinage or historical priority. The paper asks two research questions:

\begin{quote}
\textbf{RQ1.} \emph{What is Skillware, and what operational conditions distinguish it from adjacent AI-native artifacts and systems?}

\textbf{RQ2.} \emph{How does the proposed definition classify positive, negative, and boundary cases in a theoretically sampled Agent Skills ecosystem, and where does the evidence remain unresolved?}
\end{quote}

The argument begins with a programming-medium shift. Agent Skills give natural-language behavioral source persistence, identity, activation rules, support artifacts, and change history. A Skill describes reusable behavioral content. Skillware names the software boundary required to acquire, execute, maintain, and evolve that content as an independent artifact. We define an addressable Skill-centered instance of that source as a Behavioral Artifact and the independently managed software identity around one Skill or coherent Skill suite as a Skillware Unit. A compatible Agent Host activates the unit, and its Agent Runtime interprets the source with a model, context, tools, permissions, and state. Three conditions make category membership auditable. Lifecycle Continuity evaluates whether the resulting unit is managed across change as a software-grade artifact.

\begin{figure}[t]
\centering
\includegraphics[width=\linewidth]{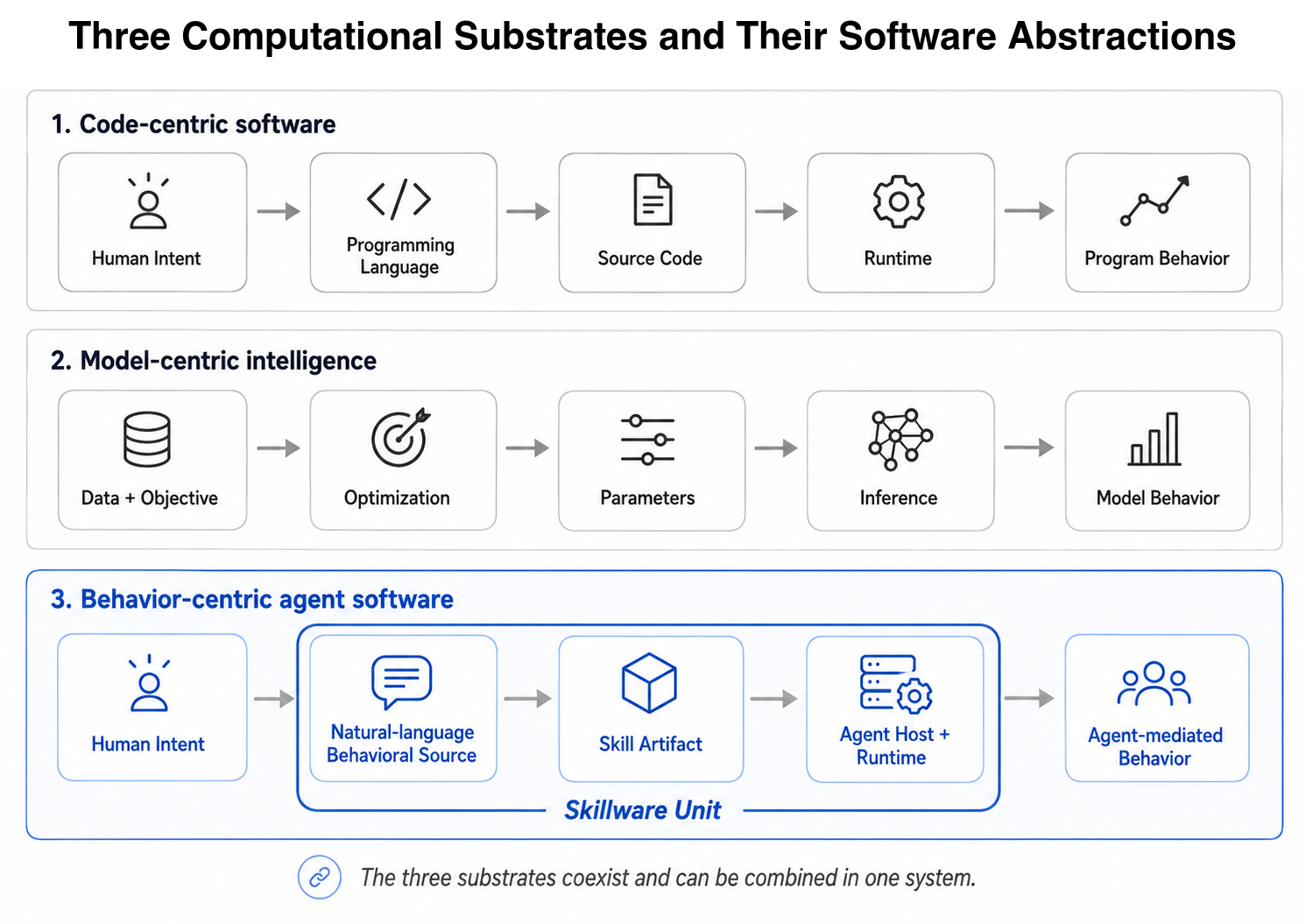}
\caption{Three computational substrates and their software abstractions. Code-, model-, and behavior-centric systems coexist; Skillware names the managed software abstraction around persistent natural-language behavioral source.}
\label{fig:programming-media}
\end{figure}

The paper makes three connected contributions. First, it defines the Skillware Ontology: a software abstraction linking Behavioral Source, Skill Artifact, Skillware Unit, Agent Host, Agent Runtime, situated performance, and task outcome. Second, it defines a Skillware Engineering Lifecycle: an initial workflow through which behavioral artifacts acquire engineering structure while remaining managed software units. Third, it introduces Identity-Preserving Evolution as a perspective on governed change in which human contributions, agent-generated proposals, and deployment evidence can produce new versions without losing software identity. Specifications, corpus signals, independent studies, and fixed-revision positive and negative cases provide the empirical basis for these claims.

The term \emph{emerging} carries a bounded meaning. Recurring artifacts and practices are already observable, while conventions, compatibility contracts, evaluation methods, and lifecycle infrastructure remain unsettled. The evidence supports existence, structural recurrence, and engineering pressure. It supplies no population estimate of high-quality Skillware, no universal adoption claim, and no behavioral-equivalence claim across agent systems.

\section{Agent Skills and the Research Landscape}

\subsection{The Existing Agent Skill Artifact}

Agent Skills provide the empirical starting point. A Skill is a directory whose required \texttt{SKILL.md} contains metadata and instructions. Optional executables, references, and assets remain inside the same directory boundary \cite{agentskills2025specification}. The loading model follows progressive disclosure: catalog metadata supports discovery, the complete instruction file enters context after activation, and supporting resources load when the workflow requires them. The implementer guide uses \emph{skills-compatible agent} for a system that discovers, activates, loads, and uses this artifact \cite{agentskills2025implementation}. We retain that established term.

The artifact already supports varied implementation carriers. An instruction-only Skill can encode a complete task contract. References and templates can supply packaged knowledge. Scripts can perform exact transformations. Hooks can bind behavior to lifecycle events. Plugins can distribute Skills with integrations. A parent Skill can select and invoke specialist Skills. A companion interface can exchange state with a running task. These carriers remain organized around the Skill when its behavioral source supplies the primary reusable task interface.

The concrete ecosystem therefore contains four interacting objects. The \emph{Skill Artifact} carries reusable behavioral source and support materials. The \emph{Skillware Unit} supplies independent software identity around one Skill or coherent suite. We use \emph{Agent Host} as an analytical role for the user-facing agent application that discovers or receives, activates, and manages the unit. An \emph{Agent Runtime}, provided or delegated by the Host, assembles context, invokes a model, routes tools, manages state, and drives an agent loop. Codex, Claude Code, GitHub Copilot, Hermes, and OpenClaw are examples when they exercise this role \cite{openai2026skills,anthropic2026skills,github2026copilotskills,hermes2026skills}. This usage is distinct from MCP's protocol-specific Host and Microsoft's AHP Agent Host process \cite{modelcontextprotocol2025architecture,microsoft2026agenthost}. The Agent Skills term \emph{skills-compatible agent} continues to describe the required discovery and activation capability \cite{agentskills2025implementation}.

\subsection{Adjacent Concepts}

Skillware intersects several established concepts. Table~\ref{tab:adjacent} distinguishes them through their primary object and their relation to a Skillware Unit. The boundaries are relational. A persistent prompt can occur inside Skillware. A plugin can package Skillware. A workflow can be encoded by Skills. An MCP server can supply tools used by Skillware. Category membership depends on C1--C3 in Section~\ref{sec:definition}; Lifecycle Continuity is evaluated separately.

\begin{table}[t]
\caption{Skillware and adjacent ecosystem concepts}
\label{tab:adjacent}
\centering
\small
\begin{tabularx}{\textwidth}{p{0.12\textwidth} p{0.19\textwidth} Y Y}
\toprule
Concept & Primary object or role & Relation to Skillware & Boundary implication \\
\midrule
Prompt & Model input or conditioning content & May carry instructions inside a Skill or an execution context & Persistence alone does not establish behavioral primacy, unit identity, and Host execution \cite{liu2023prompting} \\
Agent Skill & Addressable task-behavior module & Current primary Behavioral Artifact of Skillware & A single Skill can define the complete unit. A coherent suite can share one unit boundary \cite{agentskills2025specification} \\
Tool / MCP server & Callable operation or protocol capability & Supplies operations used during Skill execution & Tool primacy places the acquired unit in the capability layer \cite{modelcontextprotocol2025architecture} \\
Workflow & Ordering and coordination of work & Can be encoded by one Skill or composed across Skills & Workflow structure describes control flow. Skillware describes the managed software unit \\
Plugin & Packaging and integration envelope & Can contain Skills, hooks, MCP configuration, assets, and manifests & Plugin-packaged Skillware qualifies when Skills organize the reusable task behavior \cite{openai2026plugins} \\
Agent Host & User-facing application role that activates Skillware and provides or delegates runtime execution & Codex, Claude Code, GitHub Copilot, Hermes, and OpenClaw can exercise this role & Host identity remains separate from the acquired Skillware Unit \\
Agent Runtime & Context assembly, model invocation, tool routing, state, and agent-loop execution & Interprets activated Skillware inside the Host environment & Runtime configuration shapes situated performance \cite{openclaw2026runtime,openclaw2026agentloop} \\
Skillware & Software abstraction and managed unit centered on Agent Skills & Packages behavioral source, support artifacts, compatibility, and identity & Membership requires C1--C3; Lifecycle Continuity records software-grade change management \\
\bottomrule
\end{tabularx}
\end{table}

\subsection{Converging Research Lines}

Existing research can be organized by unit of analysis and engineering concern. Specification and survey work treats the Skill directory, \texttt{SKILL.md}, progressive disclosure, acquisition, and security as the main object \cite{agentskills2025specification,xu2026agentskills,zhou2026agentskills}. Runtime work treats Skills as first-class execution artifacts and identifies caching, environment construction, global management, failure handling, and security requirements. Chen et al. analyze 97,755 Skills under this framing and motivate a Skill OS abstraction \cite{skillsnewapps2026}. Their work supplies the closest application-layer and runtime precursor to our category.

Curation research examines how a Skill population changes. Ouyang et al.'s SkillOS pairs a frozen executor with a trainable curator that updates an external SkillRepo from accumulated experience \cite{ouyang2026skillos}. This architecture supports Skill selection, creation, refinement, and restructuring. Its unit of analysis is the learned curation process and repository state. The present paper absorbs it as evidence that Skills can exist as persistent external artifacts subject to experience-driven change.

An independent project from ARPA Hellenic Logical Systems also uses \emph{Skillware} for a modular framework of self-contained AI Skills \cite{arpahls2026skillware}. This work provides important prior use of the term and an implementation-level precedent in the emerging ecosystem. Its primary object is a concrete framework and package. The present paper studies a cross-project software abstraction, with an operational ontology, membership conditions, and engineering lifecycle for persistent Skill-centered artifacts. We therefore retain the distinction between prior implementation usage and the broader category definition developed here, and make no claim to word coinage or historical priority.

Repository studies establish software-engineering pressure directly. Gao et al. analyze 18,463 registry Skills and 23,199 personal-use Skills from 5,876 GitHub repositories, identify 3,709 reuse links, and code modifications after reuse \cite{registrytorepository2026}. Their findings describe copied artifacts, local bindings, additive maintenance, and comparatively stable behavioral contracts. Hong et al. study Skill anatomy and detected smells across 238 Skills, showing that Skill authoring already exhibits measurable quality problems \cite{anatomytosmells2026}. Metere treats Skills as verifiable deployment artifacts that require explicit trust and capability gates \cite{metere2026verifiableskills}.

Evolution and production research adds multi-party and governance concerns. FederatedSkill represents Skill changes as semantic diffs aggregated across clients while preserving local personalization \cite{yang2026federatedskill}. SkillFab connects demand issues, managed repositories, Git evidence, review, publication, and recovery \cite{xu2026skillfab}. These results provide concrete precedents for versioned and reviewable Skill production. They stop short of defining the cross-project software category that contains the produced artifact.

Software engineering supplies the remaining foundation. Module boundaries isolate change \cite{parnas1972modules}. Components support separate deployment and composition \cite{szyperski2002component}. Lifecycle standards organize acquisition, operation, maintenance, and disposal \cite{iso2017lifecycle}. Software-evolution research treats continuing change as intrinsic to deployed systems \cite{lehman1980evolution}. Design patterns capture recurring forces, participants, solution structures, and consequences \cite{gamma1994designpatterns,buschmann1996patternoriented,evans2003domaindriven}. The Skillware question arises where these foundations meet current Skill artifacts.

The closest studies already describe Skills as applications, engineered artifacts, verifiable artifacts, and evolving repository objects. Our bounded contribution combines these views around one declared unit. That unit must explain where reusable behavior resides, how it is acquired and executed, and how its identity survives change. The next section develops that category.

\section{From Agent Skills to Skillware}
\label{sec:definition}

\subsection{Natural-Language Behavioral Source}

Natural-language programming predates foundation models. Biermann et al. studied how users expressed procedures in natural language under constrained interpretation \cite{biermann1983natural}. Prompt programming later treated language as a way to elicit and orchestrate model behavior \cite{reynolds2021prompt}. Program synthesis uses natural language to produce code \cite{austin2021program}, while structured prompting adds control constructs and constraints around model queries \cite{beurerkellner2023prompting}. Agent Skills expose a further configuration: natural language remains present at execution time as persistent, reusable source.

We define \emph{natural-language behavioral source} as persistent linguistic instructions, constraints, policies, examples, and evaluation criteria that carry the reusable task contract of an Agent Skill. The source is human-readable and directly interpreted within an agent execution context. It can call into code and structured resources. Its active meaning depends on the current task, model, context, tools, permissions, state, and observations.

The term \emph{programming medium} captures this role precisely. Natural language supplies executable guidance without acquiring the grammar, static semantics, or determinism of a formal programming language. Formal code, learned parameters, and natural-language behavioral source can coexist. A Tool-Backed Skill, for example, can use language to decide when an exact transformation is required and a script to compute that transformation. The distinctive change lies in persistence: the linguistic source has an address, activation path, package context, and change history.

The central shift is not that natural language replaces code. Behavioral specifications themselves become persistent software artifacts with an address, execution relationship, engineering boundary, and change history. Skillware extends software engineering to this new computational object.

\subsection{Behavioral Artifact and Skillware Unit}

We operationalize a \emph{Behavioral Artifact} as an addressable artifact whose primary behavioral source specifies reusable task behavior for activation and execution by a skills-compatible agent. This definition adapts existing artifact language. Information hiding, component theory, and digital-artifact research establish independent change, separate deployment, editability, and recombination \cite{parnas1972modules,szyperski2002component,kallinikos2013digitalartifacts}. Recent Skill work already describes Skills as procedural, engineered, or verifiable artifacts \cite{registrytorepository2026,metere2026verifiableskills,zhou2026agentskills}. Our operationalization joins behavioral source to an execution path.

The relationship between the two terms is the central unit distinction: a \emph{Skill Artifact} specifies reusable behavior, while a \emph{Skillware Unit} manages that artifact as software. The Skill Artifact is the behavior carrier. The Skillware Unit is the separately addressable identity that acquires, activates, versions, maintains, and removes the carrier. One unit can contain one Skill Artifact or a coherent suite whose members share that identity.

The ontology can therefore be read as a chain of distinct responsibilities:
\begin{equation*}
\begin{aligned}
&\text{Behavioral Source} \;\rightarrow\; \text{Skill Artifact} \;\rightarrow\; \text{Skillware Unit} \\
&\qquad\xrightarrow{\text{activated by}}\; \text{Agent Host}
\;\xrightarrow{\text{interpreted by}}\; \text{Agent Runtime}
\;\rightarrow\; \text{Execution Trace} \;\rightarrow\; \text{Task Outcome}.
\end{aligned}
\end{equation*}
Behavioral Source answers where task behavior is expressed. Skill Artifact answers how that behavior is persisted. Skillware Unit answers how it becomes a software object with identity, version, provenance, and lifecycle. Agent Host answers which application activates it, and Agent Runtime answers which execution machinery interprets it. The trace and outcome remain observations of a situated run rather than properties of the source alone.

The ontology answers what the object is. A separate lifecycle axis answers how that object changes through time. Structural implementation dimensions describe the contents and operating relations of one Skillware Unit; lifecycle processes describe creation, adoption, engineering consolidation, maintenance, and evolution. A structural profile is therefore a snapshot of a unit, while a lifecycle record follows the same unit identity across releases.

Four analytical objects remain separate. The artifact specifies intended reusable behavior. A skills-compatible agent and its runtime interpret that specification. Situated performance is the resulting execution under a particular task, model, context, tool set, permission policy, and state. The task outcome is the effect evaluated against task criteria. The relation can be summarized as \emph{artifact specification $\rightarrow$ runtime interpretation $\rightarrow$ situated performance $\rightarrow$ task outcome}. This separation prevents an artifact description from serving as evidence of actual behavior or effectiveness.

A Behavioral Artifact can contain more than prose. Metadata supports discovery and compatibility. References supply domain knowledge, schemas, templates, and examples. Scripts implement deterministic kernels. Hooks connect behavior to events. Plugin manifests define a distribution boundary. Child Skills provide specialist behavior. Tests and evaluations protect contracts. Companion interfaces externalize state and human judgment. Natural-language behavioral source remains primary because it organizes how these elements participate in the task.

We define the \emph{Skillware Unit} as the independently managed software identity around one Skill Artifact or a coherent Skill suite. Name, address, package boundary, compatibility, dependencies, version, and provenance attach to this unit. In an Atomic implementation, the Skill Artifact and Skillware Unit coincide. In a compound implementation such as Superpowers, multiple specialist Skills, hooks, adapters, tests, and a Visual Companion share one package-level identity \cite{obra2026superpowers}. Update, rollback, maintenance, and removal test whether that identity also achieves Lifecycle Continuity.

The Agent Host supplies the activation boundary. It discovers or receives the Skillware Unit, determines whether its Skills are available, and hands activated behavioral source and resources to an Agent Runtime. The Runtime realizes active semantics with a task, model, assembled context, tools, permission policy, and state. This separation assigns identity to the Skillware Unit, activation to the Host, interpretation to the Runtime, situated behavior to the execution trace, and effectiveness to task evaluation.

\begin{figure}[t]
\centering
\includegraphics[width=\linewidth]{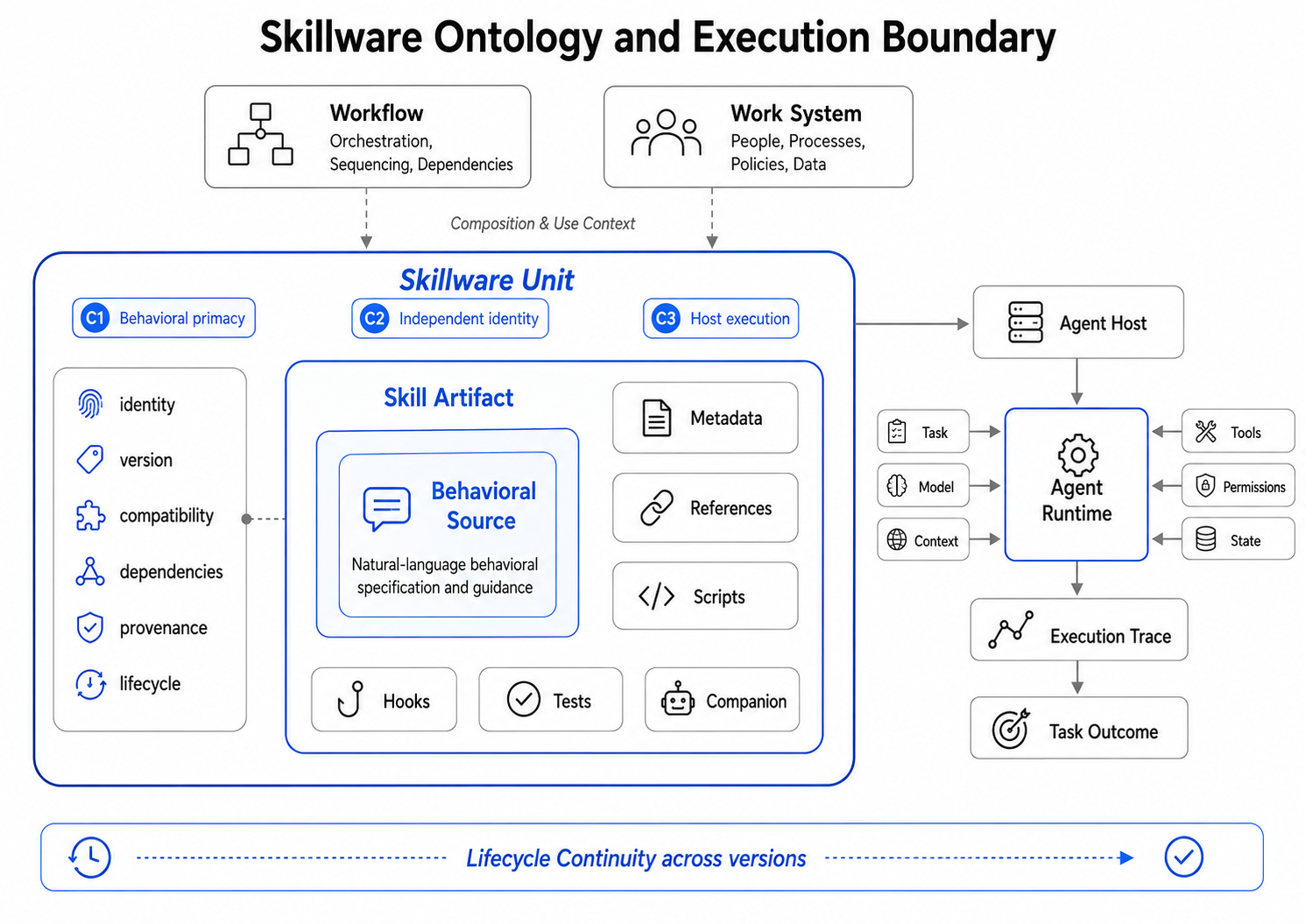}
\caption{Skillware ontology. Managed software identity, behavioral source, activation environment, runtime interpretation, situated performance, and task outcome remain distinct analytical objects.}
\label{fig:unit}
\end{figure}

\subsection{Formal Definition, Membership, and Lifecycle Grade}

\begin{center}
\fbox{\begin{minipage}{0.94\columnwidth}
\textbf{Definition 1 (Skillware).} \emph{Skillware is an emerging AI-native software abstraction whose primary Behavioral Artifact is a separately addressable Agent Skill or coherent Skill suite. A Skillware Unit is the independently managed software identity that carries this artifact as software. The Skill source carries reusable task behavior primarily through natural-language behavioral source, optionally supported by code and resources, and a compatible Agent Host can activate the unit for runtime interpretation and execution.}
\end{minipage}}
\end{center}

The definition establishes a software abstraction grounded in the current Skill ecosystem and supplies a declared evaluation unit. Three necessary conditions determine category membership. All three must point to the same Skillware Unit. Lifecycle Continuity is reported as an additional software-grade property.

The classification unit is the separately managed Skillware Unit. An individual \texttt{SKILL.md} provides evidence only within a declared unit boundary. Agent Hosts activate the unit, Agent Runtimes execute it, workflows compose its use, and organizational work systems can depend on it. These surrounding systems retain their own identities. For this paper, membership is assigned only when C1, C2, and C3 hold for the same unit. Lifecycle Continuity is recorded as an orthogonal software-grade attribute. Skillware Units can later be assembled into portfolios or embedded in larger work systems while remaining the lifecycle objects evaluated by this definition.

In this definition, \emph{AI-native} identifies software whose primary task behavior depends on model-mediated interpretation during execution. Conventional code, schemas, data, services, and user interfaces can implement bounded responsibilities inside the unit. The Skill-centered behavioral source continues to organize how those components participate in the task.

\textbf{C1: Skill-centered behavioral primacy.} One Skill or a coherent Skill suite carries and organizes the reusable task behavior. The Skill source defines the task interface, decision procedure, constraints, and relation to support components. Supporting code can be substantial. The category test asks which artifact remains primary in the acquired unit's behavior, routing, and user-facing identity.

\textbf{C2: Independent software identity.} The candidate unit has an addressable software identity that can be acquired, installed, and tracked independently of the Agent Host. A directory, package, installer, plugin, or repository release can establish observable name, address, version, compatibility, and provenance records.

\textbf{C3: Agent Host execution relationship.} At least one identified compatible Agent Host can discover or receive, activate, load, and use the unit through skills-compatible operations. A documented or reproduced path in Codex, Claude Code, GitHub Copilot, Hermes, OpenClaw, or another compatible system is sufficient for category membership. Cross-Host parity remains a separate compatibility question.

\textbf{Lifecycle Continuity (software-grade property).} The same unit identity persists across update, maintenance, disablement or rollback, and removal. Responsibility can be distributed among the project, installer, package manager, lifecycle infrastructure, and Agent Host. Lifecycle Continuity measures the strength of software-grade management after C1--C3 establish category membership.

\begin{table}[t]
\caption{Operational tests for Skillware membership and software-grade lifecycle}
\label{tab:conditions}
\centering
\footnotesize
\begin{tabularx}{\columnwidth}{p{0.10\columnwidth} p{0.15\columnwidth} Y Y}
\toprule
Test & Role & Positive observable & Frequent exclusion signal \\
\midrule
C1 & Membership & Skill source organizes the reusable task contract and support artifacts & Skill content is incidental to a tool, index, or application \\
C2 & Membership & Stable, independently tracked software identity with address, package, version, compatibility, or provenance & Instructions remain session-bound or inseparable from the containing system \\
C3 & Membership & Identified Agent Host discovery, activation, loading, and use path & Artifact is documentation or catalog content only \\
LC & Software grade & Update, maintenance, rollback, and removal preserve the same unit identity & Change replaces the containing system or loses unit identity \\
\bottomrule
\end{tabularx}
\end{table}

Membership under C1--C3 does not score quality, security, usefulness, popularity, implementation complexity, or lifecycle maturity. A compact single-file Skill can satisfy the category conditions. A large plugin can fall outside the category when application code or callable services carry the primary task behavior. Lifecycle Continuity and other engineering properties are evaluated after category membership is established.

\subsection{Execution Semantics and Runtime Responsibility}

The active semantics of Skillware are context-relative. Let $S$ denote the Skillware Unit, $H$ the Agent Host, $\tau$ the task, $M$ the model, $C$ the assembled context, $T$ the available tools, $P$ the permission policy, and $\sigma$ the current state. The Host activates $S$, and its provided or delegated Agent Runtime $R_H$ produces an execution trace $\pi$:

\begin{equation}
\pi = R_H(S, \tau, M, C, T, P, \sigma).
\label{eq:runtime}
\end{equation}

The trace can include text, tool calls, files, state transitions, approvals, review stops, and failures. The Host controls availability and activation; the Runtime loads Skill source and resources, invokes model inference, routes allowed tools, incorporates observations, updates state, and continues or stops the agent loop \cite{agentskills2025implementation,yao2023react,openclaw2026runtime,openclaw2026agentloop}. Equation~\ref{eq:runtime} assigns no deterministic semantics to unrestricted natural language. It records the variables that a compatibility or robustness claim must control.

The closest conventional analogy combines interpreter and runtime responsibilities. The Agent Runtime interprets Skill source in a task context, and model inference participates in that interpretation. Some actions perform compilation-like lowering into plans, tool arguments, shell commands, or generated artifacts. The paper introduces no generic compiler component for Agent systems. Established runtime and agent-loop vocabulary provides the more defensible abstraction.

This responsibility split improves failure analysis. A failure can originate in ambiguous Skill source, missing context, model variation, unavailable tools, permission policy, agent-system-specific activation, stale dependencies, or corrupted state. A Skillware release should therefore record its artifact version together with relevant execution configuration. Artifact quality and runtime quality remain connected and separately measurable.

\subsection{Boundary Cases}

Negative cases clarify the unit. An index of external Skills supports discovery while carrying no primary Skill behavior of its own. A reference MCP filesystem server is independently installable and callable, while tools constitute its primary artifact. An agent-system-internal configuration can be edited and deleted without a separately acquired Skill-centered boundary. A plugin with one incidental Skill remains plugin software when executable services or application code define the primary behavior.

Positive boundary cases span a wide range. A single installable \texttt{SKILL.md} can satisfy C1--C3. A Skill plus references and scripts can form one Resource-Backed and Tool-Backed unit. A coherent suite can include multiple Skills, activation hooks, system adapters, tests, release metadata, and a companion while retaining Skill primacy. Lifecycle Continuity can differ across these members. These cases show why directory names and file counts cannot classify the category. The primary behavioral artifact, independent unit identity, and Agent Host execution relationship must converge.

\section{Open-Source Evidence}

\subsection{Evidence Design}

The empirical argument uses four evidence layers with separate inference boundaries. First, the Agent Skills specification and official implementation documentation establish the artifact envelope, progressive disclosure, and named system behavior. Second, a frozen metadata corpus measures scale, frontmatter delimiters, document size, repository identifiers, and explicit references to packaged support paths. Third, fixed-revision case review establishes unit boundaries and operational relations. Fourth, independent empirical studies contribute reuse, maintenance, quality, verification, and runtime findings under their own samples and denominators.

Category review begins by declaring the candidate unit and freezing its revision. The review then records positive evidence, counterevidence, and limitations for C1--C3, with Lifecycle Continuity coded separately. Inclusion requires affirmative support for the complete C1--C3 conjunction. Missing or contradictory membership evidence leaves the relevant condition unresolved and supplies no basis for inclusion. This protocol keeps classification attached to one named unit and makes disagreements traceable to evidence paths or boundary choices.

This separation prevents the corpus from carrying claims it cannot support. A \texttt{SKILL.md} path token does not prove that a resource exists or participates in execution. A repository directory named \texttt{hooks} does not prove event coupling. A long instruction file does not prove low quality. Collection-scale signals motivate engineering questions. Operational claims require specifications, executable paths, tests, or fixed-revision source evidence.

The public evidence supplement contains three linked sets \cite{fan2026skillwarepatterns}. The SkillMD-138K snapshot provides collection-scale metadata \cite{skillmd138k}. A category-boundary review contains fifteen declared units, twelve inclusions and three deliberate negatives, assessed against C1--C3 with lifecycle evidence reported separately. A technical review contains thirteen fixed-revision implementation units selected across Skill-unit size, packaged resources, executable support, runtime-event coupling, plugin packaging, companion interaction, and adaptation mechanisms. The units are Superpowers, gstack, ECC, last30days, scientific-schematics, financial-analysis, dot-skill, ui-ux-pro-max, Caveman, design-taste-frontend, darwin-skill, SkillOpt-Sleep, and OpenMontage. Each technical case records a canonical repository, 40-character commit, exact paths, positive observations, counterevidence, and limitations.

\begin{figure}[t]
\centering
\includegraphics[width=\linewidth]{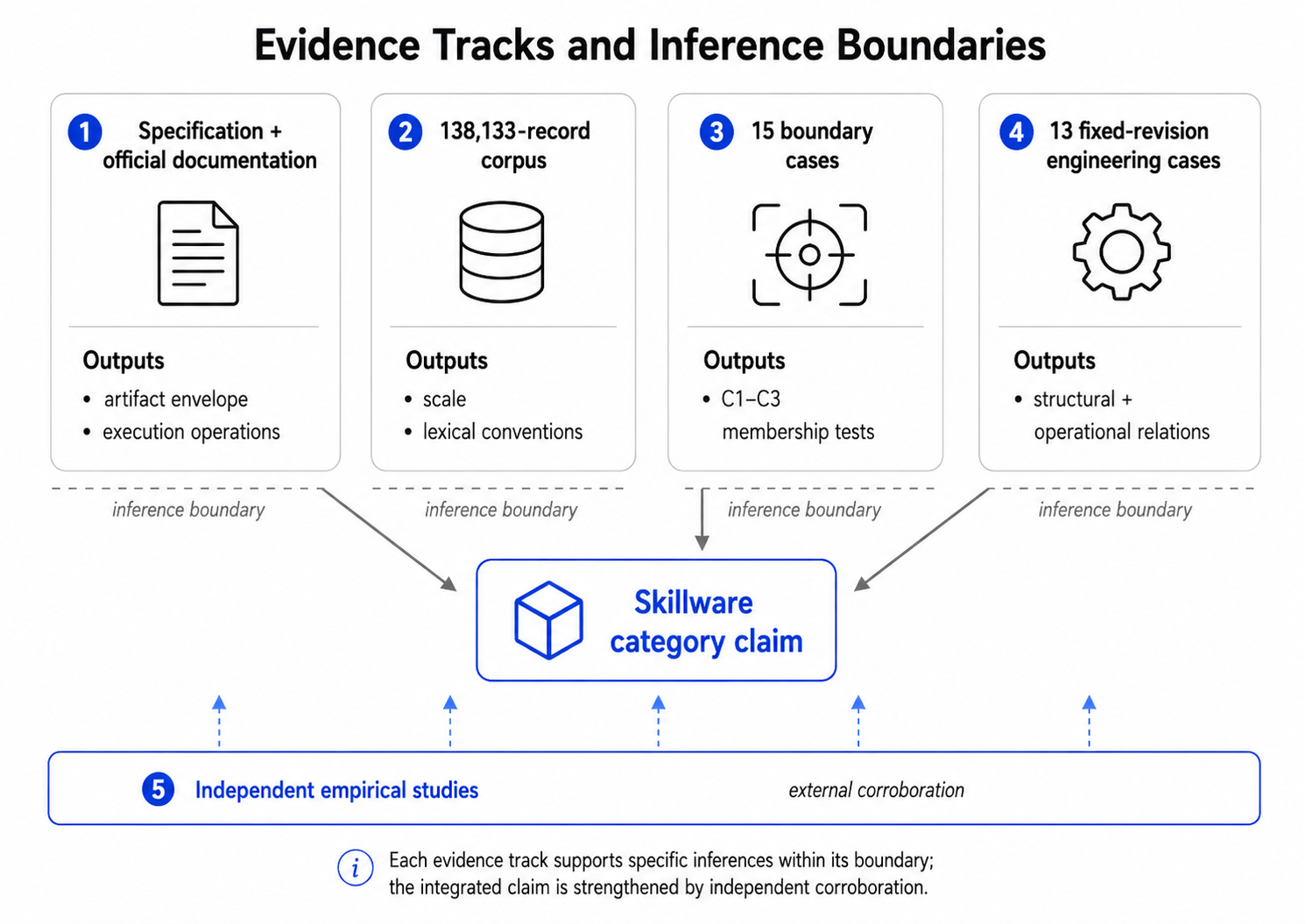}
\caption{Evidence design. Corpus, case, specification, and independent-study results retain distinct units and claim boundaries.}
\label{fig:evidence}
\end{figure}

\subsection{Collection-Scale Signals}

The analyzed SkillMD-138K snapshot is fixed at revision \texttt{0d73048a\allowbreak bf2fb6ee\allowbreak 91f6f9f5\allowbreak ac598d5b\allowbreak e8d6bdd7}. Its raw Parquet object has SHA-256 \texttt{c3c07af0\allowbreak 19bc36fb\allowbreak 0ffdfaef\allowbreak 46df52da\allowbreak ffb66ca7\allowbreak 9422990a\allowbreak 74c58e41\allowbreak 6d1e6403}. Content-hash deduplication yields 138,133 \texttt{SKILL.md} records associated with 20,556 repository identifiers. Frontmatter delimiters appear in 136,380 files, or 98.73\% of the corpus. This measurement supports one narrow conclusion: a structured metadata envelope is a dominant visible convention in the collected artifacts. It does not validate frontmatter fields or establish identical activation semantics across agent systems. The public evidence supplement publishes the metadata manifest, aggregate outputs, repository aliases, verification records, and frozen case files while excluding third-party Skill bodies \cite{fan2026skillwarepatterns}.

A narrow lexical detector finds 32,069 files, approximately 23.2\%, with at least one explicit path token into \texttt{references/}, \texttt{scripts/}, or \texttt{assets/}. Across the corpus, it finds 236,421 such tokens. These results show that explicit references to adjacent artifacts form a measurable authoring convention. The detector misses implicit dependencies, command names, remote services, hooks, generated adapters, and files selected through undocumented conventions. It can also count illustrative or stale paths. Repository-level resolution is required for compound-anatomy claims.

Document length supplies an additional engineering signal. The median file has 169 lines and 687 words, while the 90th percentile has 517 lines and 1,971 words. Length carries no direct quality or maturity meaning. It indicates the amount of behavioral source that may enter discovery, context assembly, review, and maintenance. Large instruction surfaces motivate explicit task contracts, progressive disclosure, resource modularization, and regression evaluation. Controlled studies must test whether those mechanisms actually improve outcomes.

The value of the corpus lies in scale, recurrence, and question formation. It shows that the observed phenomenon extends across many collected artifacts and that structured metadata plus explicit support-path references recur often enough to demand software-engineering analysis. Pattern occurrence, runtime reliability, usefulness, security, and production adoption remain outside the corpus inference boundary.

\subsection{Independent Empirical Evidence}

Independent studies strengthen the engineering interpretation with separate methods. Chen et al. characterize 97,755 Skills and identify locality, semi-deterministic blocks, environment dependencies, caching, global management, failure handling, and security as system requirements \cite{skillsnewapps2026}. Their application analogy and Skill OS proposal support the premise that Skill execution has become a systems problem.

Gao et al. show that Skills are authored, copied, customized, and maintained across registry and personal repositories \cite{registrytorepository2026}. Their study reports 18,463 registry Skills, 23,199 personal-use Skills from 5,876 GitHub repositories, and 3,709 reuse links. It also identifies operational specifications, local bindings, and additive domain knowledge as important change surfaces. Hong et al. analyze 238 Skills and report widespread detected smells \cite{anatomytosmells2026}. These results establish artifact-level maintenance and quality pressure while retaining their source-specific samples and detector assumptions.

Verification, curation, and production work fill other parts of the lifecycle. Skills as Verifiable Artifacts supplies explicit trust-state and capability-gate requirements \cite{metere2026verifiableskills}. SkillOS uses accumulated experience to update an external SkillRepo \cite{ouyang2026skillos}. SkillFab ties demand, Git history, review, registry publication, and recovery into one production system \cite{xu2026skillfab}. Together these studies show that Skill identity now participates in execution, evaluation, change, and governance.

\begin{table}[t]
\caption{Evidence-to-condition mapping and inference limits}
\label{tab:evidence-map}
\centering
\small
\begin{tabularx}{\textwidth}{p{0.13\textwidth} Y Y Y}
\toprule
Condition & Specification and corpus evidence & Case and independent-study evidence & Inference limit \\
\midrule
C1 Primacy & Required \texttt{SKILL.md}, instructions, and optional support directories & Skill-centered packages, reuse, and maintained behavioral contracts & Lexical structure cannot establish primacy by itself \\
C2 Identity & Directory format, name, metadata, compatibility fields & Stable repository paths, manifests, installers, versions, and provenance & A path does not prove independent acquisition or identity \\
C3 Host execution & Progressive disclosure and skills-compatible implementation operations & Host bindings, hooks, scripts, companion round trips, runtime studies, and tests & Source inspection does not establish cross-Host behavioral parity \\
LC Software grade & Version-capable metadata and managed Skill scopes & Release records, updates, maintenance, verification, recovery, and removal paths & Current public cases provide uneven lifecycle coverage \\
\bottomrule
\end{tabularx}
\end{table}

\subsection{Fixed-Revision Cases}

The technical cases reveal how repository structure maps to proposed operating relations. Superpowers contains a coherent Skill suite, a SessionStart bootstrap, agent-system-specific package manifests, tests, scripts, and a brainstorming Visual Companion that exchanges task state with a browser \cite{obra2026superpowers}. gstack combines a Skill-based entry surface with generated system bindings, executable services, state, hooks, and a browser round trip \cite{garrytan2026gstack}. ECC packages Skills, Agents, commands, rules, hooks, and system-specific configuration \cite{affaan2026ecc}. OpenMontage exposes staged Skill composition, checkpoints, lifecycle events, and a companion workflow \cite{calesthio2026openmontage}. SkillOpt supplies execution-evidence optimization, staging, validation, backup, and adoption mechanisms for a target Skill \cite{microsoft2026skillopt}.

Negative cases test the same unit logic. An external Skill index lacks primary behavioral content. The MCP filesystem reference server centers callable tools. An agent-system-managed GPT configuration lacks the separately acquired Agent Skill boundary required by the current definition. These exclusions show that installability, executability, and persistence can exist outside Skillware. Behavioral primacy, independent identity, and the Agent Host execution relationship perform the classification.

This first application establishes discriminability within the reviewed set. It supplies no estimate of inter-rater reliability. Independent coding and agreement analysis remain required to test whether third parties reproduce the same boundary decisions.

The evidence supports the emergence claim. A stable artifact envelope exists. Many collected files explicitly reference supporting artifacts. Independent studies observe execution requirements, reuse, maintenance, and quality pressure. Fixed projects exhibit independent package identities, Host bindings, event, tool, companion, release, and partial adaptation surfaces. Lifecycle Continuity varies across the reviewed units, which supports treating it as a measurable software-grade property. Production reliability and ecosystem prevalence require additional measurements.

\section{Skillware as a Software Engineering Object}

\subsection{Engineering Continuity}

If Skillware is a software abstraction for persistent Behavioral Artifacts, established software-engineering concepts should remain applicable to its managed units. We define \emph{Skillware Engineering} as the construction, evaluation, packaging, release, maintenance, and evolution of Skillware Units. Module boundaries guide separation of behavioral responsibilities \cite{parnas1972modules}. Component theory guides unit identity, interfaces, dependencies, deployment, and composition \cite{szyperski2002component}. Lifecycle standards guide acquisition, operation, maintenance, and retirement \cite{iso2017lifecycle}. Skill-specific studies add empirical evidence of reuse, local adaptation, quality smells, and verification requirements \cite{registrytorepository2026,anatomytosmells2026,metere2026verifiableskills}. The following structures and pattern mappings test this continuity through observable participant relations.

Skillware also creates specific obligations. Natural-language source can be ambiguous and sensitive to surrounding context. Model and runtime changes can alter execution. Tool calls create side effects. Hooks bind behavior to system events. Plugins introduce compatibility and supply-chain surfaces. Companions create task-state and authorization problems. Adaptation mechanisms can optimize against incomplete evidence. The software category makes these obligations assignable to a managed unit and version.

\subsection{Recurring Implementation Dimensions of Skillware Units}

We identify one minimal baseline and six recurring implementation dimensions inside declared Skillware Units. Atomic is the exclusive minimal profile. Resource-Backed, Tool-Backed, Event-Driven, Plugin-Packaged, Companion-Coupled, and Adaptation-Enabled describe independent relations that can compose into a multi-label profile. These labels summarize operating surfaces. They carry no maturity, quality, or implementation-history semantics.

These dimensions belong to the structural axis. They answer what is inside a Skillware Unit at a declared revision. The lifecycle axis remains independent: a small Atomic unit can be maintained for years, while a highly packaged unit can be newly created. No structural label implies a lifecycle stage or a required development order.

\begin{figure}[t]
\centering
\includegraphics[width=\linewidth]{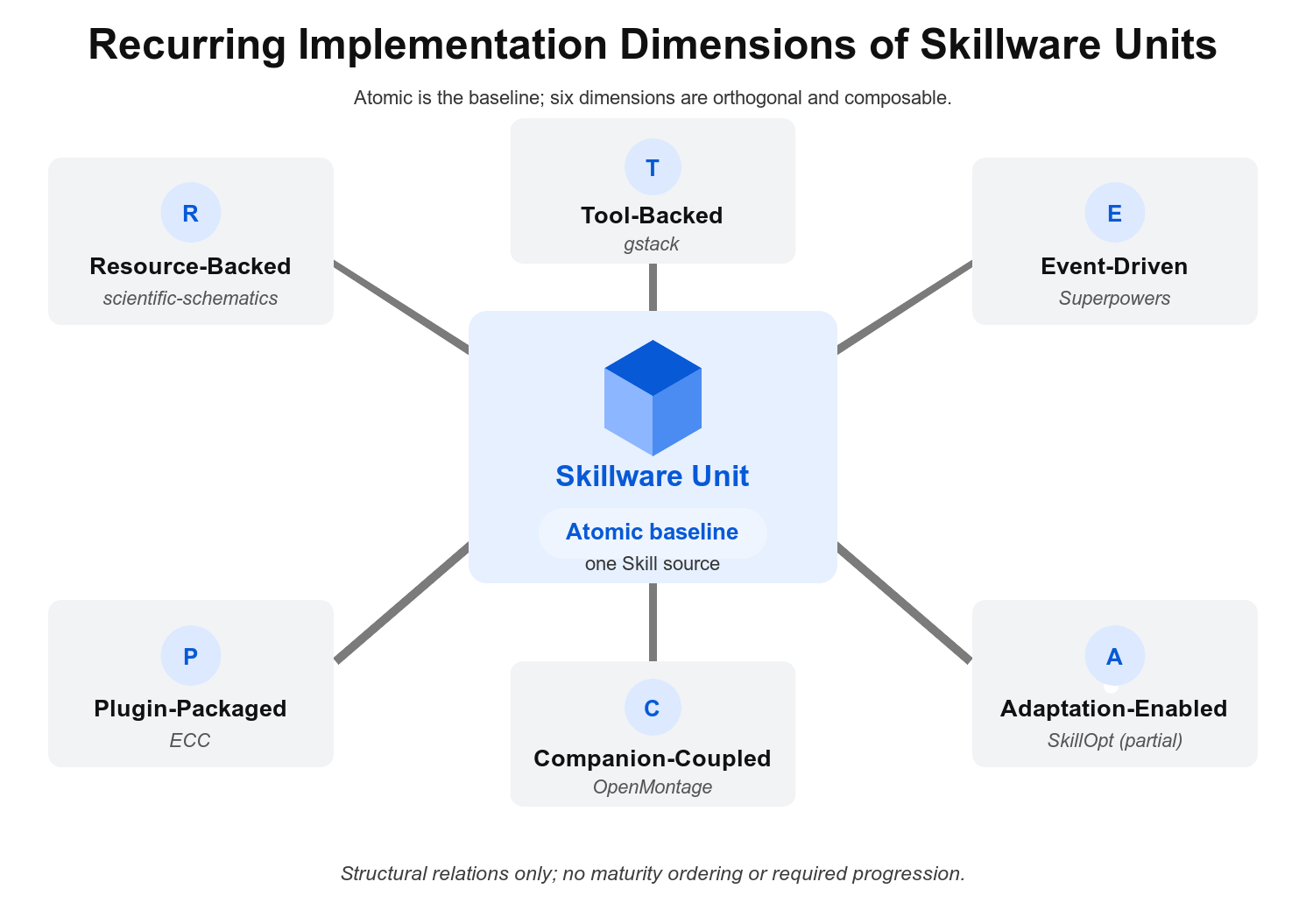}
\caption{Atomic is the baseline configuration of a Skillware Unit. Six orthogonal and composable dimensions describe additional implementation relations without maturity ordering.}
\label{fig:forms}
\end{figure}

\begin{table}[t]
\caption{Implementation dimensions, qualifying relations, risks, and examples}
\label{tab:forms}
\centering
\footnotesize
\begin{tabularx}{\textwidth}{p{0.14\textwidth} Y Y p{0.20\textwidth}}
\toprule
Profile or dimension & Qualifying relation & Principal engineering risk & Reviewed example \\
\midrule
Atomic & One Skill source contains the evaluated task contract with no required packaged support component & Hidden ambiguity and monolithic instruction growth & design-taste-frontend \\
Resource-Backed & The Skill loads packaged references, templates, data, examples, or assets & Missing, stale, conflicting, or excessive context & scientific-schematics, financial-analysis \\
Tool-Backed & The workflow invokes a bundled script, executable, daemon, service, or tool & Input contracts, side effects, dependency failure, and error propagation & gstack, OpenMontage \\
Event-Driven & A concrete lifecycle event activates, initializes, constrains, or observes behavior & Duplicate, missing, reordered, or non-idempotent events & Superpowers, ECC \\
Plugin-Packaged & A system-recognized manifest distributes the Skillware Unit and integrations & Compatibility, provenance, upgrade, and uninstall drift & Superpowers, ECC \\
Companion-Coupled & A separate interface completes a task-state round trip & Session binding, stale decisions, authorization, and recovery & Superpowers Visual Companion, gstack browser \\
Adaptation-Enabled & A governed mechanism can propose, evaluate, adopt, release, and recover changes to the same unit identity & Evaluation contamination, unsafe adoption, provenance loss, and rollback failure & No complete case among the 13 reviewed units. SkillOpt supplies partial Self-Evolution evidence \\
\bottomrule
\end{tabularx}
\end{table}

The profile rule preserves mixed structures. Superpowers is Resource-Backed, Tool-Backed, Event-Driven, Plugin-Packaged, and Companion-Coupled in the frozen review. gstack is Resource-Backed, Tool-Backed, Event-Driven, and Companion-Coupled. OpenMontage combines resource, tool, event, and companion relations. The profile states which surfaces must be engineered and tested. A profile with more labels receives no automatic quality advantage.

\subsection{Design Pattern Transfer as Software-Continuity Evidence}

Design patterns capture recurring software forces rather than implementation syntax. A successful transfer from conventional software to Skillware tests whether responsibilities, interfaces, contracts, and dependencies can be constructed when the carrier changes \cite{gamma1994designpatterns}. The main text focuses on Facade, Adapter, Composite, Observer, State, and Strategy. Together they exercise encapsulation, interface adaptation, composition, event notification, state-dependent behavior, and interchangeable decision procedures. The repository supplement at \url{https://github.com/MetaInFLow/skillware-patterns} screens all 23 Gang of Four patterns and also examines Pipes and Filters and Specification from their established architectural and domain traditions \cite{buschmann1996patternoriented,evans2003domaindriven,fan2026skillwarepatterns}. The source patterns remain prior software-engineering knowledge. The paper contributes participant mappings grounded in Skillware implementations.

Pattern implementation is independent of file type. A Facade can be an entry Skill backed by a bootstrap and invocation policy. An Adapter can use generated Host bindings. A Composite can join parent and child Skills under one stage contract. An Observer can use hooks and event records. State can use persisted checkpoints to constrain permitted transitions. Scripts, services, manifests, large plugins, and child Skills are valid carriers when the participant relation spans the declared Skillware Unit.

We admit a transfer when seven elements are available: source intent, design forces, participant correspondence, consequences, implementation evidence, focused verification, and a misuse discriminator. This protocol rejects mappings based solely on names. A file called \texttt{adapter.yaml} is only navigational evidence until it translates a canonical interface into a target contract. A router becomes Facade when it gives clients a unified access contract over an independently addressable subsystem.

\subsubsection{Superpowers \texttt{using-superpowers} as Facade}

Superpowers provides the principal open-source mapping. The evaluated unit contains specialist workflow Skills and a \texttt{using-superpowers} Skill that establishes one access policy for discovering and invoking them. Task-level agent execution is the Client. \texttt{using-superpowers} is the Facade. The specialist Skills form the subsystem. The Agent Host's skills-compatible interface supplies the Skill-loading operation. A SessionStart hook bootstraps the Facade into the session, while the selection and invocation policy implements its operation \cite{obra2026superpowers}.

The frozen Skill states the access rule compactly:

\begin{lstlisting}
If you think there is even a 1% chance a skill might apply to what you are doing,
you ABSOLUTELY MUST invoke the skill.
\end{lstlisting}

This excerpt shows a declarative Facade operation. It requires the Client to enter the subsystem through one policy and keeps specialist Skills independently addressable. The Facade is therefore implemented through natural-language behavioral source, bootstrap, and the agent system's invocation primitive. A large plugin method is unnecessary for this occurrence. Other pattern occurrences can use scripts, services, manifests, or child Skills.

\subsubsection{Canonical Host Bindings as Adapter}

An Adapter separates a canonical component contract from a target-system contract. This force appears when one Skillware Unit must be activated through different Host discovery paths, command names, or field conventions. gstack provides the open-source correspondence through generated and system-specific bindings \cite{garrytan2026gstack}. The repository's canonical-decision-brief fixture states the invariant directly:

\begin{lstlisting}
Accept canonical fields `topic` and `decision`; produce `brief_path`.
Host bindings may translate discovery paths, command names, and field names,
but must not change this procedure.
\end{lstlisting}

The canonical Skill is the Adaptee, the Host binding is the Adapter, and the skills-compatible activation path is the Target interface. The mapping is useful because Host variation becomes an explicit compatibility boundary. Runtime parity still requires tests under each target Host.

\subsubsection{Skill Suites as Composite}

A Composite gives clients one contract for both an atomic component and a structured group of components. OpenMontage's staged Skill workflows provide the ecosystem correspondence \cite{calesthio2026openmontage}. The constructive fixture makes the containment rule explicit:

\begin{lstlisting}
Every component accepts `request` and emits one `section` record.
A leaf runs its atomic procedure. A composite runs children from
`workflow.yaml`, rejects cycles, and combines their section records.
\end{lstlisting}

The client invokes one Skill-stage contract. A leaf Skill or a parent Skill suite implements the same result shape, while the workflow file supplies part--whole composition. This is software composition: the suite has a shared identity and contract, while the internal Skills remain independently inspectable.

\subsubsection{Lifecycle Hooks as Observer}

Observer separates a subject's state transition from independent consumers of the resulting event. ECC hooks and OpenMontage lifecycle events supply open-source correspondences \cite{affaan2026ecc,calesthio2026openmontage}. The release-subject fixture uses a typed event and explicit delivery accounting:

\begin{lstlisting}
After a successful release transition, publish one `brief.released.v1` event.
Notify only registered observers. Record delivery per observer,
isolate failures, and prevent re-entrant release.
\end{lstlisting}

The Skillware Unit is the Subject, the lifecycle hook is the notification point, and registered downstream Skills or services are Observers. The mapping exposes event ordering, registration, failure isolation, and re-entrancy as testable obligations.

\subsubsection{Persisted Checkpoints as State}

State makes permitted behavior depend on an explicit state representation and transition relation. OpenMontage checkpoints and SkillOpt stages provide partial ecosystem correspondences \cite{calesthio2026openmontage,microsoft2026skillopt}. The stateful-workflow fixture binds action selection to persisted state:

\begin{lstlisting}
Load the current state before acting. Apply only a transition declared in
`states.yaml`, atomically persist the target state, and reject illegal
and stale transitions. On restart, recover the persisted state.
\end{lstlisting}

The state object is the persisted checkpoint, transitions are the allowed Skill operations, and recovery is the restart path. The mapping shows how a natural-language procedure can carry state invariants while executable persistence supplies atomicity.

\subsubsection{Routing Policies as Strategy}

Strategy represents interchangeable decision procedures behind one request contract. A Skill router or a selection policy can choose among specialist Skills while preserving the client's task interface. Superpowers' selection and invocation policy provides the open-source motivation \cite{obra2026superpowers}. The evidence-strategy fixture expresses the common contract and explicit fallback:

\begin{lstlisting}
Select `fast-scan` for low-risk reversible decisions and `deep-review` otherwise.
Each strategy accepts the same evidence request and returns `claims`, `sources`,
and `confidence`. Use the declared fallback for unknown selection.
\end{lstlisting}

The Strategy interface is the shared evidence request and result contract; the concrete strategies are interchangeable procedures. This occurrence shows that natural-language source can carry a decision policy with explicit selection forces and compatibility constraints.

These transfers do not replace conventional code. They show that software architecture is defined by responsibilities, forces, and contracts rather than by the medium that carries behavior. Natural-language source, scripts, hooks, plugins, and child Skills can each implement a participant when the declared Skillware Unit preserves the pattern's structure and consequences.

\begin{table}[t]
\caption{Representative design-pattern transfers and open-source correspondences}
\label{tab:patterns}
\centering
\footnotesize
\begin{tabularx}{\textwidth}{p{0.11\textwidth} p{0.22\textwidth} Y Y}
\toprule
Pattern & Skillware participant relation & Open-source correspondence & Claim status \\
\midrule
Facade & Entry Skill exposes one stable access contract over specialist Skills & Superpowers \texttt{using-superpowers}, gstack root routing & Superpowers participant mapping verified in the frozen case \\
Adapter & Thin binding translates canonical Skill semantics into a target system contract & gstack generated bindings and system-specific surfaces & Strong correspondence. Parity requires runtime tests \\
Composite & Atomic and composite stages share one invocation and result contract & OpenMontage staged Skill workflows & Candidate correspondence plus constructive repository fixture \\
Observer & Subject emits typed events to registered independent consumers & ECC hooks, OpenMontage lifecycle events & Candidate pending complete registration and delivery evidence \\
State & Persisted state controls permitted actions and transitions & OpenMontage checkpoints, SkillOpt stages & Strong structural candidate \\
Strategy & Stable request and result contract selects among interchangeable procedures & Superpowers selection policy and evidence-strategy fixture & Constructive mapping; comparative benefit requires runtime study \\
\bottomrule
\end{tabularx}
\end{table}

The full screen is retained in the repository supplement. Ten GoF patterns currently have detailed constructive implementations: Facade, Adapter, Composite, Decorator, Strategy, Template Method, Observer, State, Memento, and Mediator. Pipes and Filters and Specification add two patterns from other traditions. Each example includes a real \texttt{SKILL.md}, participant map, mapping-specific artifact, structurally close misuse, and focused tests. The six mappings in Table~\ref{tab:patterns} provide the main-text continuity argument. The supplement supports constructibility across a broader pattern set. Ecosystem frequency and comparative benefit remain empirical questions.

\subsection{Engineering Mechanisms and Method}

Patterns and mechanisms serve different decisions. A transferred pattern preserves an established participant structure. An engineering mechanism is a recurring implementation practice that can realize several patterns or connect them. The current catalog contains eight mechanisms: Skill Router, Progressive Disclosure, Deterministic Kernel, Canonical Skill with Generated System Adapters, Lifecycle Bootstrap, Verification Gate, Human-Gated Companion, and Evidence-Gated Self-Evolution.

Skill Router centralizes predicates, precedence, targets, and fallback. Progressive Disclosure binds resources to explicit loading conditions. Deterministic Kernel moves exact transformations behind versioned executable contracts. Generated adapters keep one canonical behavioral source while translating thin system bindings. Lifecycle Bootstrap attaches idempotent initialization to a bounded event. Verification Gate binds consequential transitions to fresh evidence. Human-Gated Companion connects one paused task to a validated human decision. Evidence-Gated Self-Evolution stages candidate changes, evaluates them, and preserves release identity and rollback.

As an initial workflow derived from the software abstraction, we organize these decisions into six steps:

\begin{enumerate}
\item \textbf{Define the Skillware Unit and task contract.} Record the unit identity, triggering conditions, inputs, outputs, invariants, permissions, side effects, failure behavior, and lifecycle owner.
\item \textbf{Select implementation dimensions from operating relations.} Mark the resources, tools, events, packages, companions, and adaptation surfaces required by the task.
\item \textbf{Select patterns from design forces.} Map clients, components, events, state, strategies, snapshots, or mediators using the source pattern's participant contract.
\item \textbf{Implement engineering mechanisms.} Add deterministic boundaries, loading conditions, system bindings, lifecycle hooks, verification, human gates, and governed change where the forces require them.
\item \textbf{Evaluate and package the complete unit.} Test behavioral contracts, exact operations, negative cases, system compatibility, failure recovery, upgrade, and removal against the candidate release.
\item \textbf{Release, observe, maintain, and adapt.} Preserve provenance, execution evidence, version history, review decisions, deprecation policy, and rollback paths.
\end{enumerate}

The workflow keeps three decision layers explicit. Implementation dimensions identify operating surfaces. Patterns organize participant relations. Mechanisms realize behavior. Tests and evidence validate the combined release. This ordering also makes design alternatives comparable: two implementations can share one dimensional profile while using different patterns or mechanisms.

\begin{figure}[t]
\centering
\includegraphics[width=\linewidth]{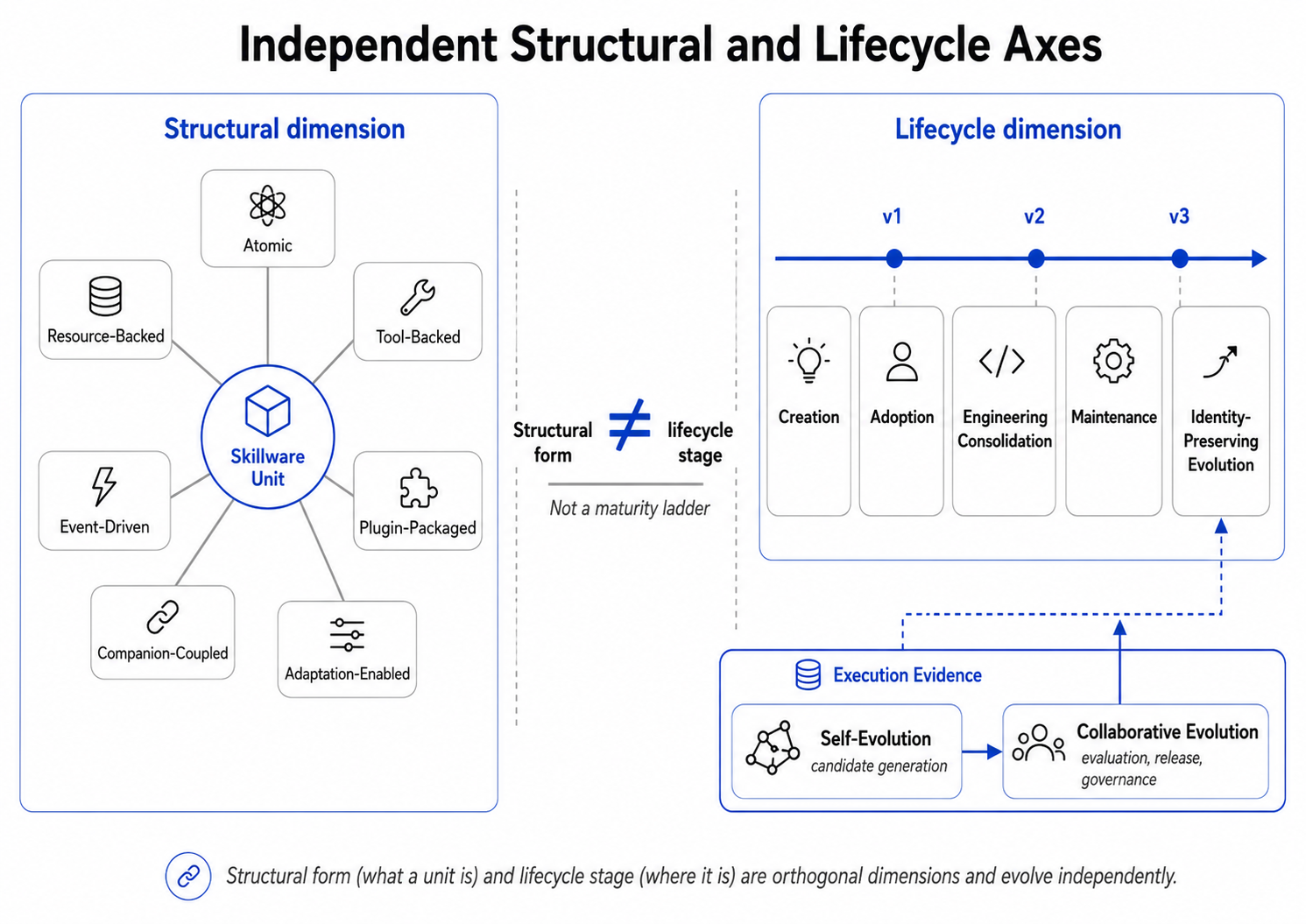}
\caption{Structural and lifecycle dimensions of Skillware. A unit's implementation profile is a snapshot; identity-preserving lifecycle change follows the same managed software object through versions.}
\label{fig:evolution}
\end{figure}

\section{Evolution of Behavioral Software Artifacts}

\subsection{Skillware Development Trajectory and Engineering Consolidation}

Natural-language behavioral source supports rapid implementation because a useful task contract can be authored before every recurring operation has a stable code representation. A possible Skillware development trajectory therefore begins with a natural-language behavioral seed, becomes a reusable Skill, acquires a Skillware Unit identity, and accumulates engineering structure as the unit is used. We define \emph{Engineering Consolidation} as the evidence-driven process through which a natural-language-first Skillware Unit gains explicit resources, deterministic scripts, system bindings, packages, companions, tests, provenance, and lifecycle controls across releases.

This trajectory is a possible development path, not a maturity ladder or a required progression. A unit may remain compact and instruction-centered, or it may become a hybrid software system with substantial executable and interface support. The continuity anchor is the managed behavioral artifact and its unit identity, rather than the amount of engineering structure accumulated around it.

\begin{figure}[t]
\centering
\includegraphics[width=\linewidth]{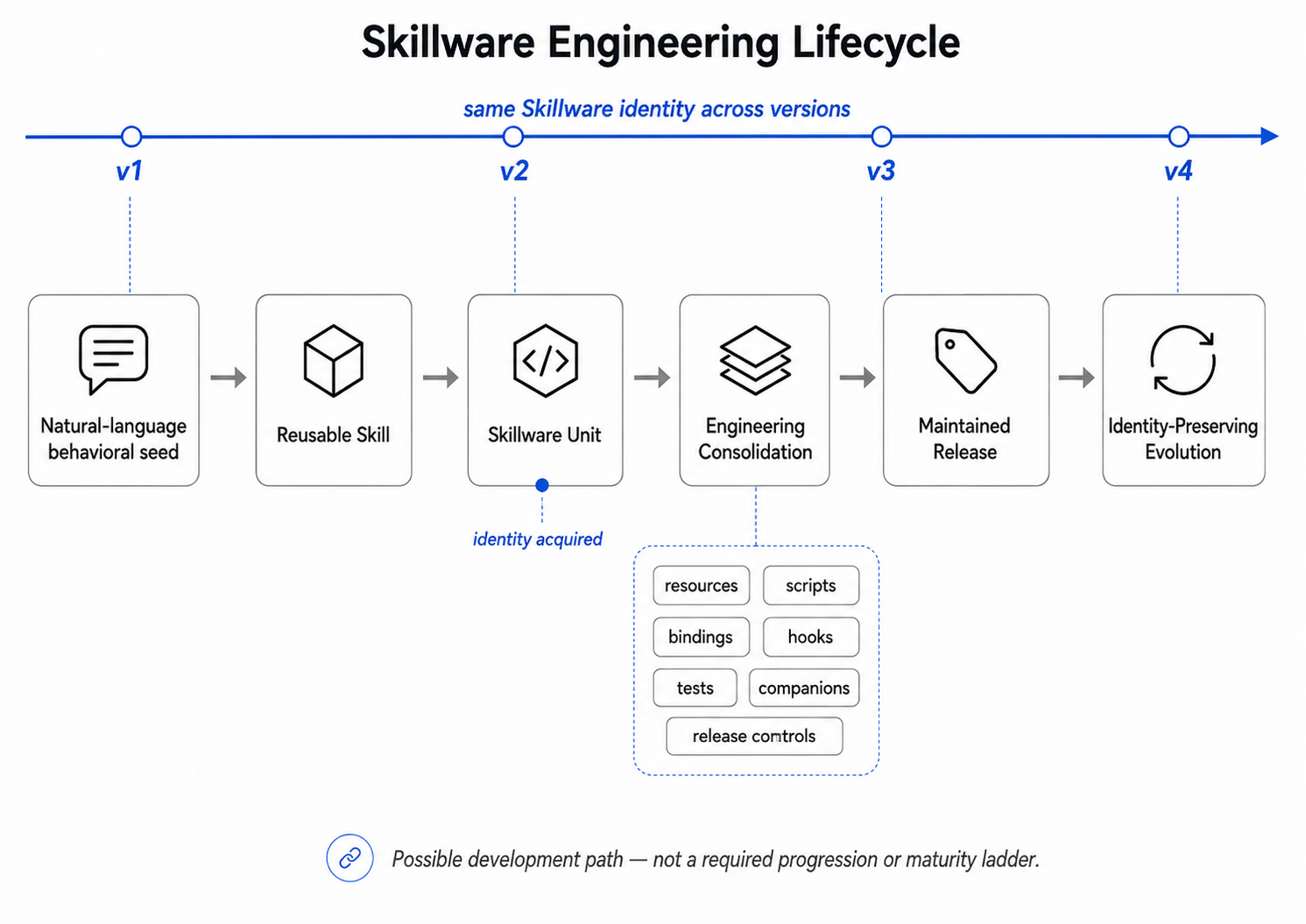}
\caption{A possible Skillware engineering lifecycle. The path organizes creation, adoption, evidence-driven consolidation, maintenance, and governed evolution around one continuing Skillware Unit; it is neither a maturity ladder nor a required sequence.}
\label{fig:engineering-lifecycle}
\end{figure}

Existing evidence supports the existence of software-like behavioral artifacts and engineering pressures. The longitudinal transition from initial Skill creation to mature Skillware remains an empirical research question requiring release- and commit-level studies.

Consolidation is independent of implementation-dimension labels. A release may add a deterministic validator and become Tool-Backed. Another release may remove a companion while strengthening tests. A Resource-Backed unit can refine schemas without changing its dimensional profile. The governing input is operating evidence: stability, safety, reuse frequency, cost, failure impact, context pressure, and observability determine where additional engineering structure has value. The original behavioral source may become increasingly encapsulated by these components while the Skillware identity remains the continuity anchor.

Existing evidence supports the components of this process. Gao et al. observe changes to operational specifications, local bindings, and accumulated domain knowledge \cite{registrytorepository2026}. Chen et al. identify recurring deterministic blocks and environment dependencies \cite{skillsnewapps2026}. Superpowers and gstack show hybrid Skill-centered systems with hooks, services, tests, adapters, and companions \cite{obra2026superpowers,garrytan2026gstack}. SkillFab shows demand-driven creation, Git evidence, review, publication, and recovery \cite{xu2026skillfab}.

\subsection{Identity-Preserving Evolution}

Conventional software evolution moves from requirements and operating evidence through developer-authored changes, review, and release. Skillware adds execution traces and model-assisted proposals as direct inputs to the change process. We define \emph{Identity-Preserving Evolution} as a change process in which a Skillware Unit receives distributed contributions while retaining its software identity, provenance, compatibility, and lifecycle history. A Skillware evolution event occurs when an accepted change produces a new version of the same managed Behavioral Artifact. Within-run adaptation, model updates, prompt-context variation, and unadopted candidate patches remain separate phenomena.

The lifecycle change can be stated compactly:
\begin{equation*}
\text{execution evidence} \;\rightarrow\; \text{behavior-change proposal}
\;\rightarrow\; \text{evaluation} \;\rightarrow\; \text{governed release}.
\end{equation*}
The proposal may be generated by a human, an agent, or a system operating on execution evidence. The software lifecycle begins at release only after evaluation, decision authority, provenance, and recovery obligations are explicit. The identity-preserving path is analogous to issue, proposal, evaluation, review, merge, and release in conventional software maintenance: the version changes while the managed unit remains the same software object.

\subsubsection{Self-Evolution and Collaborative Evolution}

Self-improving systems provide the lineage for automated change. The Goedel Machine formalizes self-modification justified against an objective, and the Darwin Goedel Machine studies empirical open-ended improvement across agent variants \cite{schmidhuber2007godel,zhang2026darwin}. SkillOS learns a curation policy that changes an external SkillRepo from experience \cite{ouyang2026skillos}. SkillOpt harvests execution evidence, proposes candidate changes, validates them, and exposes staging, backup, and adoption surfaces \cite{microsoft2026skillopt}.

These systems commonly produce a candidate patch, revised target Skill, semantic diff, or updated Skill repository. In the Skillware lifecycle, Self-Evolution names this candidate-generation path. Adoption requires evaluation, review authority, version assignment, provenance, release, rejection handling, and recovery under the same Skillware Unit identity. None of the thirteen frozen technical units satisfied every requirement. SkillOpt and darwin-skill remain partial mechanism precedents.

We use \emph{Collaborative Evolution} for one possible realization of Identity-Preserving Evolution. It describes the governance topology through which multiple humans, agents, field engineers, and deployments contribute evidence or change proposals. It has adjacent foundations in open-source governance, FederatedSkill, and SkillFab \cite{mockus2002opensource,omahony2007governance,yang2026federatedskill,xu2026skillfab}. We propose a future Skillware workflow whose output is a reviewed and versioned shared release:

\begin{equation}
\begin{split}
&\text{field evidence and proposals} \rightarrow \text{evaluation} \rightarrow \text{review} \\
&\rightarrow \text{release} \rightarrow \text{redistribution} \rightarrow \text{observation and rollback}.
\end{split}
\end{equation}

Self-Evolution and Collaborative Evolution therefore occupy different roles. Self-Evolution supplies candidate changes from execution evidence. Collaborative Evolution is one governance realization that evaluates, integrates, and releases contributions across participants. Review authority, release identity, compatibility declarations, provenance, rejection, and rollback remain explicit. This workflow is future work. Federated semantic patches and reviewable Skill production provide partial precedent, while an end-to-end governed shared-release study remains open.

\subsection{Deployment and Field Adaptation Hypothesis}

Deployment and field adaptation offer one possible implication of behavioral-artifact evolution. In a Forward Deployed Engineering (FDE) setting, a field team can deploy a natural-language-first task contract, inspect its behavior with domain users, collect failures and recurring needs, and consolidate stable behavior into resources, scripts, bindings, tests, and releases. A central team can preserve architecture, review standards, compatibility, and shared versions while local teams retain site-specific adaptations. This is a delivery hypothesis about one operating context, rather than a claim about organizational structure.

The expected advantages are faster initial delivery, a readable behavioral interface, and direct conversion of field evidence into versioned artifacts. The risks include local overfitting, weak evaluation, fragmented variants, privacy leakage, and review overload. Future field studies should compare Skillware-centered and code-first delivery on time to first useful deployment, reuse across sites, defect rate, review effort, consolidation quality, regression frequency, and propagation time for accepted releases.

\section{Discussion, Limitations, and Conclusion}

\subsection{What Changes When Behavior Becomes an Artifact}

The Skillware abstraction clarifies what the Agent era adds to software composition. Formal code remains essential, and learned parameters remain central to model behavior. Persistent natural-language behavioral source adds a software surface that developers can read, distribute, activate, evaluate, and change. Agent Skills give that source a concrete carrier. The Skillware Unit makes independent software identity explicit around the carrier, while Lifecycle Continuity measures how that identity survives change.

This framing preserves the ecosystem relation. Skills are the primary Behavioral Artifacts inside Skillware. The Skillware Unit carries independent identity and engineering obligations. The Agent Host activates the unit. The Agent Runtime realizes active semantics through model inference, context, tools, permissions, state, and an agent loop. Workflows and organizational systems compose the resulting capability without becoming the classification unit.

The programming-medium claim has a bounded scope. Natural language provides behavioral source under Agent-mediated interpretation. It lacks the stable formal semantics of a conventional programming language. Equation~\ref{eq:runtime} makes this dependence visible. Behavioral contracts, evaluation suites, compatibility records, and execution traces become essential because the same artifact can vary across models, runtime versions, contexts, and tool availability.

Design-pattern transfer provides a strong test of software-engineering continuity. The value comes from participant structure and forces. Superpowers' Facade mapping identifies a Client, a unified Skill entry contract, independently addressable specialist Skills, and a bootstrap plus invocation mechanism. Adapter, Composite, Observer, State, and Strategy mappings create similarly testable relations. These mappings demonstrate that selected software-engineering participant relations can be constructed across text, scripts, plugins, events, child Skills, and companions. This analysis treats those carriers as valid only when they satisfy the declared pattern contract.

\subsection{Threats to Validity}

The collection-scale evidence is metadata- and lexical-signal based. Registry-heavy collection, aliases, forks, generated artifacts, missing source revisions, and unequal repository contribution constrain external validity. The released manifest and provenance records make the obtained snapshot auditable, while content-level replication requires access to the separately acquired corpus. Frontmatter and path-token measurements cannot establish semantic correctness or operational use.

The category-boundary and engineering cases are purposively selected. They clarify variation and counterexamples without estimating ecosystem prevalence. Fixed commits improve reproducibility and age as projects change. Source inspection establishes documented relations at those revisions. Direct multi-system execution is required for reliability and parity claims. Several close studies are recent preprints whose methods and conclusions may change through review.

Construct validity also remains open. Skill-centered primacy can be difficult to judge in packages with substantial executable services. Different analysts may select different unit boundaries or Agent Host roles. The three membership conditions, separate Lifecycle Continuity property, counterevidence fields, and frozen paths make disagreements inspectable. Independent coding, carrier-diverse negative cases, and inter-rater analysis should validate the definition.

The implementation dimensions and pattern mappings are analytical constructs. A multi-label profile compresses architectural detail. Constructive examples establish implementability and misuse discrimination while leaving prevalence and benefit unresolved. No complete Adaptation-Enabled Skillware case was established among the thirteen frozen technical units. Collaborative Evolution and FDE suitability remain prospective claims.

\subsection{Research Agenda}

The immediate agenda concerns robustness. Studies should compare alternative Skillware implementations under fixed tasks, Agent Hosts, models, Agent Runtime versions, tool availability, dependencies, permissions, and sampling settings. Repeated paired trials should report task success, constraint violations, between-run variance, latency, cost, recovery, and human-review burden. Dimensional profiles identify comparable surfaces. Patterns and mechanisms identify the design variables.

Behavioral contracts and versioning form a second agenda. Research is needed on invariant representation, semantic versioning for natural-language behavioral source, compatibility declarations, system adapters, and tests that separate artifact quality from model and runtime effects. Security work should cover provenance, capability gates, package installation, prompt injection through references, command and path safety, companion sessions, adaptive supply chains, and rollback.

Longitudinal studies should follow Engineering Consolidation across releases. They can measure which instructions move into resources or deterministic kernels, which integrations become system bindings, and how test coverage and maintenance cost change. Evolution studies should measure candidate quality, held-out improvement, gate selectivity, provenance completeness, shared-review workload, version propagation, and recovery under noisy or adversarial feedback.

\subsection{Conclusion}

Agent Skills show that reusable task behavior can be carried primarily by persistent natural-language source and packaged with code, resources, events, integrations, tests, and interfaces. This paper defines that source as natural-language behavioral source, operationalizes the Skill as a Behavioral Artifact, and defines Skillware as the AI-native software abstraction built around independently managed Agent Skills.

Three necessary conditions make category membership auditable: Skill-centered behavioral primacy, independent software identity, and an Agent Host execution relationship. Lifecycle Continuity records software-grade management across update, maintenance, rollback, and removal. The Agent Skills specification, a 138,133-record corpus, independent studies, negative cases, and fixed-revision implementations provide converging evidence that the form already exists and faces software-engineering pressure.

Skillware also extends an engineering tradition. One minimal profile and six recurring implementation dimensions expose structural operating surfaces. Representative design patterns organize participants across text, scripts, plugins, events, child Skills, and companions. The Skillware Engineering Lifecycle connects the task contract to evaluation, release, maintenance, and adaptation. Engineering Consolidation describes how field evidence can add explicit engineering structure across versions. Identity-Preserving Evolution describes how candidate changes can become new versions without losing the managed unit identity; Collaborative Evolution is one possible realization of that process.

The central claim is compact: persistent natural-language behavioral specifications create a managed software surface in agent systems. Skillware is the software abstraction that extends software engineering to this surface by providing addressable artifacts, independent identity, documented or reconstructed activation relations, engineering structure, and a basis for identity-preserving change.

\bibliographystyle{IEEEtran}
\bibliography{ref}

@misc{fan2026skillwarepatterns,
  author = {Fan, Anthony and Lan, Neil},
  title = {Skillware Patterns: Executable, Bilingual Pattern-Transfer Records for Skillware},
  year = {2026},
  howpublished = {GitHub repository, release v0.1-paper-v1},
  url = {https://github.com/MetaInFLow/skillware-patterns/releases/tag/v0.1-paper-v1},
  note = {Paper-bound release at commit cf328ba6ccaf3102cb3857bc5d88990bd161ba91}
}

@article{parnas1972modules,
  author = {Parnas, David L.},
  title = {On the Criteria To Be Used in Decomposing Systems into Modules},
  journal = {Communications of the ACM},
  volume = {15},
  number = {12},
  pages = {1053--1058},
  year = {1972},
  doi = {10.1145/361598.361623},
  url = {https://doi.org/10.1145/361598.361623}
}

@article{rumelhart1986backprop,
  author = {Rumelhart, David E. and Hinton, Geoffrey E. and Williams, Ronald J.},
  title = {Learning Representations by Back-Propagating Errors},
  journal = {Nature},
  volume = {323},
  pages = {533--536},
  year = {1986},
  doi = {10.1038/323533a0},
  url = {https://doi.org/10.1038/323533a0}
}

@article{lecun2015deep,
  author = {LeCun, Yann and Bengio, Yoshua and Hinton, Geoffrey},
  title = {Deep Learning},
  journal = {Nature},
  volume = {521},
  pages = {436--444},
  year = {2015},
  doi = {10.1038/nature14539},
  url = {https://doi.org/10.1038/nature14539}
}

@article{biermann1983natural,
  author = {Biermann, Alan W. and Ballard, Bruce W. and Sigmon, Anne H.},
  title = {An Experimental Study of Natural Language Programming},
  journal = {International Journal of Man-Machine Studies},
  volume = {18},
  number = {1},
  pages = {71--87},
  year = {1983},
  doi = {10.1016/S0020-7373(83)80005-4},
  url = {https://doi.org/10.1016/S0020-7373(83)80005-4}
}

@inproceedings{reynolds2021prompt,
  author = {Reynolds, Laria and McDonell, Kyle},
  title = {Prompt Programming for Large Language Models: Beyond the Few-Shot Paradigm},
  booktitle = {Extended Abstracts of the 2021 CHI Conference on Human Factors in Computing Systems},
  pages = {1--7},
  year = {2021},
  doi = {10.1145/3411763.3451760},
  url = {https://doi.org/10.1145/3411763.3451760}
}

@article{austin2021program,
  author = {Austin, Jacob and Odena, Augustus and Nye, Maxwell and Bosma, Maarten and Michalewski, Henryk and Dohan, David and Jiang, Ellen and Cai, Carrie and Terry, Michael and Le, Quoc V. and Sutton, Charles},
  title = {Program Synthesis with Large Language Models},
  journal = {arXiv preprint arXiv:2108.07732},
  year = {2021},
  eprint = {2108.07732},
  archivePrefix = {arXiv},
  doi = {10.48550/arXiv.2108.07732},
  url = {https://arxiv.org/abs/2108.07732}
}

@article{beurerkellner2023prompting,
  author = {Beurer-Kellner, Luca and Fischer, Marc and Vechev, Martin},
  title = {Prompting Is Programming: A Query Language for Large Language Models},
  journal = {Proceedings of the ACM on Programming Languages},
  volume = {7},
  number = {PLDI},
  pages = {1946--1969},
  year = {2023},
  doi = {10.1145/3591300},
  url = {https://doi.org/10.1145/3591300}
}

@article{lehman1980evolution,
  author = {Lehman, Meir M.},
  title = {Programs, Life Cycles, and Laws of Software Evolution},
  journal = {Proceedings of the IEEE},
  volume = {68},
  number = {9},
  pages = {1060--1076},
  year = {1980},
  doi = {10.1109/PROC.1980.11805},
  url = {https://doi.org/10.1109/PROC.1980.11805}
}

@incollection{schmidhuber2007godel,
  author = {Schmidhuber, J{\"u}rgen},
  title = {G{\"o}del Machines: Fully Self-Referential Optimal Universal Self-Improvers},
  booktitle = {Artificial General Intelligence},
  series = {Cognitive Technologies},
  pages = {199--226},
  publisher = {Springer},
  year = {2007},
  doi = {10.1007/978-3-540-68677-4_7},
  url = {https://doi.org/10.1007/978-3-540-68677-4_7}
}

@inproceedings{zhang2026darwin,
  author = {Zhang, Jenny and Hu, Shengran and Lu, Cong and Lange, Robert Tjarko and Clune, Jeff},
  title = {Darwin G{\"o}del Machine: Open-Ended Evolution of Self-Improving Agents},
  booktitle = {International Conference on Learning Representations},
  year = {2026},
  eprint = {2505.22954},
  archivePrefix = {arXiv},
  doi = {10.48550/arXiv.2505.22954},
  url = {https://openreview.net/forum?id=pUpzQZTvGY}
}

@article{mockus2002opensource,
  author = {Mockus, Audris and Fielding, Roy T. and Herbsleb, James D.},
  title = {Two Case Studies of Open Source Software Development: Apache and Mozilla},
  journal = {ACM Transactions on Software Engineering and Methodology},
  volume = {11},
  number = {3},
  pages = {309--346},
  year = {2002},
  doi = {10.1145/567793.567795},
  url = {https://doi.org/10.1145/567793.567795}
}

@article{omahony2007governance,
  author = {O'Mahony, Siobh{\'a}n and Ferraro, Fabrizio},
  title = {The Emergence of Governance in an Open Source Community},
  journal = {Academy of Management Journal},
  volume = {50},
  number = {5},
  pages = {1079--1106},
  year = {2007},
  doi = {10.5465/amj.2007.27169153},
  url = {https://doi.org/10.5465/amj.2007.27169153}
}

@article{bommasani2021foundation,
  author = {Bommasani, Rishi and Hudson, Drew A. and Adeli, Ehsan and others},
  title = {On the Opportunities and Risks of Foundation Models},
  journal = {arXiv preprint arXiv:2108.07258},
  year = {2021},
  eprint = {2108.07258},
  archivePrefix = {arXiv},
  doi = {10.48550/arXiv.2108.07258},
  url = {https://arxiv.org/abs/2108.07258}
}

@article{ouyang2026skillos,
  author = {Ouyang, Siru and Yan, Jun and Chen, Yanfei and Han, Rujun and Wang, Zifeng and others},
  title = {{SkillOS}: Learning Skill Curation for Self-Evolving Agents},
  journal = {arXiv preprint arXiv:2605.06614},
  year = {2026},
  eprint = {2605.06614},
  archivePrefix = {arXiv},
  doi = {10.48550/arXiv.2605.06614},
  url = {https://arxiv.org/abs/2605.06614}
}

@book{szyperski2002component,
  author = {Szyperski, Clemens and Gruntz, Dominik and Murer, Stephan},
  title = {Component Software: Beyond Object-Oriented Programming},
  edition = {2},
  publisher = {Addison-Wesley Professional},
  year = {2002},
  isbn = {9780201745726}
}

@inproceedings{yao2023react,
  author = {Yao, Shunyu and Zhao, Jeffrey and Yu, Dian and Du, Nan and Shafran, Izhak and Narasimhan, Karthik and Cao, Yuan},
  title = {{ReAct}: Synergizing Reasoning and Acting in Language Models},
  booktitle = {International Conference on Learning Representations},
  year = {2023},
  url = {https://openreview.net/forum?id=WE_vluYUL-X}
}

@article{liu2023prompting,
  author = {Liu, Pengfei and Yuan, Weizhe and Fu, Jinlan and Jiang, Zhengbao and Hayashi, Hiroaki and Neubig, Graham},
  title = {Pre-train, Prompt, and Predict: A Systematic Survey of Prompting Methods in Natural Language Processing},
  journal = {ACM Computing Surveys},
  volume = {55},
  number = {9},
  pages = {1--35},
  year = {2023},
  doi = {10.1145/3560815},
  url = {https://doi.org/10.1145/3560815}
}

@misc{iso2017lifecycle,
  author = {{ISO/IEC/IEEE}},
  title = {{ISO/IEC/IEEE 12207:2017}: Systems and Software Engineering---Software Life Cycle Processes},
  year = {2017},
  url = {https://www.iso.org/standard/63712.html},
  urldate = {2026-07-21}
}

@misc{agentskills2025specification,
  author = {{Agent Skills}},
  title = {Agent Skills Specification},
  year = {2025},
  url = {https://agentskills.io/specification},
  urldate = {2026-07-21}
}

@misc{agentskills2025implementation,
  author = {{Agent Skills}},
  title = {How to Add Skills Support to Your Agent},
  year = {2025},
  url = {https://agentskills.io/client-implementation/adding-skills-support},
  urldate = {2026-07-21}
}

@misc{openai2026skills,
  author = {{OpenAI}},
  title = {Build Skills},
  year = {2026},
  url = {https://learn.chatgpt.com/docs/build-skills},
  urldate = {2026-07-21}
}

@misc{openai2026plugins,
  author = {{OpenAI}},
  title = {Build Plugins},
  year = {2026},
  url = {https://learn.chatgpt.com/docs/build-plugins},
  urldate = {2026-07-21}
}

@misc{anthropic2026skills,
  author = {{Anthropic}},
  title = {Agent Skills in the Claude Agent SDK},
  year = {2026},
  url = {https://code.claude.com/docs/en/agent-sdk/skills},
  urldate = {2026-07-21}
}

@misc{github2026copilotskills,
  author = {{GitHub}},
  title = {About Agent Skills},
  year = {2026},
  url = {https://docs.github.com/en/copilot/concepts/agents/about-agent-skills},
  urldate = {2026-07-21}
}

@misc{modelcontextprotocol2025architecture,
  author = {{Model Context Protocol}},
  title = {Architecture},
  year = {2025},
  url = {https://modelcontextprotocol.io/specification/2025-06-18/architecture},
  urldate = {2026-07-21}
}

@book{gamma1994designpatterns,
  author = {Gamma, Erich and Helm, Richard and Johnson, Ralph and Vlissides, John},
  title = {Design Patterns: Elements of Reusable Object-Oriented Software},
  publisher = {Addison-Wesley},
  address = {Reading, MA},
  year = {1994},
  isbn = {9780201633610}
}

@book{buschmann1996patternoriented,
  author = {Buschmann, Frank and Meunier, Regine and Rohnert, Hans and Sommerlad, Peter and Stal, Michael},
  title = {Pattern-Oriented Software Architecture, Volume 1: A System of Patterns},
  publisher = {John Wiley \& Sons},
  address = {Chichester, UK},
  year = {1996}
}

@book{evans2003domaindriven,
  author = {Evans, Eric},
  title = {Domain-Driven Design: Tackling Complexity in the Heart of Software},
  publisher = {Addison-Wesley},
  address = {Boston, MA},
  year = {2003}
}

@article{registrytorepository2026,
  author = {Gao, Haoyu and Lulla, Jai Lal and Lin, Hong Yi and Baltes, Sebastian and Treude, Christoph and Zahedi, Mansooreh},
  title = {From Registry to Repository: How AI Agent Skills Are Written, Adapted, and Maintained},
  journal = {arXiv preprint arXiv:2607.00911},
  year = {2026},
  eprint = {2607.00911},
  archivePrefix = {arXiv},
  doi = {10.48550/arXiv.2607.00911},
  url = {https://arxiv.org/abs/2607.00911}
}

@article{anatomytosmells2026,
  author = {Hong, David Boram and Imani, Aaron and Ahmed, Iftekhar},
  title = {From Anatomy to Smells: An Empirical Study of {SKILL.md} in Agent Skills},
  journal = {arXiv preprint arXiv:2607.01456},
  year = {2026},
  eprint = {2607.01456},
  archivePrefix = {arXiv},
  doi = {10.48550/arXiv.2607.01456},
  url = {https://arxiv.org/abs/2607.01456}
}

@misc{skillmd138k,
  author = {{FayeZC}},
  title = {{SkillMD-138K}: A Large Public Collection of Agent Skill Files},
  year = {2026},
  howpublished = {Hugging Face dataset},
  url = {https://huggingface.co/datasets/FayeZC/SkillMD-138K},
  note = {Snapshot revision 0d73048abf2fb6ee91f6f9f5ac598d5be8d6bdd7}
}

@misc{obra2026superpowers,
  author = {Vincent, Jesse and contributors},
  title = {Superpowers},
  year = {2026},
  howpublished = {GitHub repository},
  url = {https://github.com/obra/superpowers/tree/896224c4b1879920ab573417e68fd51d2ccc9072},
  note = {Evaluated at fixed revision 896224c4b1879920ab573417e68fd51d2ccc9072}
}

@misc{garrytan2026gstack,
  author = {Tan, Garry and contributors},
  title = {gstack},
  year = {2026},
  howpublished = {GitHub repository},
  url = {https://github.com/garrytan/gstack/tree/11de390be1be6849eb9a15f91ff4922dd16c589a},
  note = {Evaluated at fixed revision 11de390be1be6849eb9a15f91ff4922dd16c589a}
}

@misc{affaan2026ecc,
  author = {{ECC contributors}},
  title = {{ECC}},
  year = {2026},
  howpublished = {GitHub repository},
  url = {https://github.com/affaan-m/ECC/tree/2bc924faf2f8e893bfe0af86b1931283693c30ae},
  note = {Evaluated at fixed revision 2bc924faf2f8e893bfe0af86b1931283693c30ae}
}

@misc{calesthio2026openmontage,
  author = {{calesthio}},
  title = {{OpenMontage}},
  year = {2026},
  howpublished = {GitHub repository},
  url = {https://github.com/calesthio/OpenMontage/tree/db91727598d08d40919d7d68a47864a5467bd448},
  note = {Evaluated at fixed revision db91727598d08d40919d7d68a47864a5467bd448}
}

@misc{microsoft2026skillopt,
  author = {{Microsoft}},
  title = {{SkillOpt}: Optimizing Agent Skills from Execution Evidence},
  year = {2026},
  howpublished = {GitHub repository},
  url = {https://github.com/microsoft/SkillOpt/tree/b860a5cf88ce75e2bd02ca981ac21fb28cffba83},
  note = {Evaluated at fixed revision b860a5cf88ce75e2bd02ca981ac21fb28cffba83}
}

@article{kallinikos2013digitalartifacts,
  author = {Kallinikos, Jannis and Aaltonen, Aleksi and Marton, Attila},
  title = {The Ambivalent Ontology of Digital Artifacts},
  journal = {MIS Quarterly},
  volume = {37},
  number = {2},
  pages = {357--370},
  year = {2013},
  doi = {10.25300/MISQ/2013/37.2.02},
  url = {https://doi.org/10.25300/MISQ/2013/37.2.02}
}

@article{metere2026verifiableskills,
  author = {Metere, Alfredo},
  title = {Skills as Verifiable Artifacts: A Trust Schema and a Biconditional Correctness Criterion for Human-in-the-Loop Agent Runtimes},
  journal = {arXiv preprint arXiv:2605.00424},
  year = {2026},
  eprint = {2605.00424},
  archivePrefix = {arXiv},
  doi = {10.48550/arXiv.2605.00424},
  url = {https://arxiv.org/abs/2605.00424}
}

@article{xu2026agentskills,
  author = {Xu, Renjun and Yan, Yang},
  title = {Agent Skills for Large Language Models: Architecture, Acquisition, Security, and the Path Forward},
  journal = {arXiv preprint arXiv:2602.12430},
  year = {2026},
  eprint = {2602.12430},
  archivePrefix = {arXiv},
  doi = {10.48550/arXiv.2602.12430},
  url = {https://arxiv.org/abs/2602.12430}
}

@article{zhou2026agentskills,
  author = {Zhou, Yingli and Shu, Wang and Su, Yaodong and Du, Wenchuan and Fang, Yixiang and Lin, Xuemin},
  title = {A Comprehensive Survey on Agent Skills: Taxonomy, Techniques, and Applications},
  journal = {arXiv preprint arXiv:2605.07358},
  year = {2026},
  eprint = {2605.07358},
  archivePrefix = {arXiv},
  doi = {10.48550/arXiv.2605.07358},
  url = {https://arxiv.org/abs/2605.07358}
}

@misc{microsoft2026agenthost,
  author = {{Microsoft}},
  title = {{VS Code} Agent Host Architecture},
  year = {2026},
  url = {https://code.visualstudio.com/docs/agents/concepts/agent-host},
  urldate = {2026-07-21}
}

@misc{openclaw2026runtime,
  author = {{OpenClaw contributors}},
  title = {Agent Runtime Architecture},
  year = {2026},
  url = {https://docs.openclaw.ai/agent-runtime-architecture},
  urldate = {2026-07-21}
}

@misc{openclaw2026agentloop,
  author = {{OpenClaw contributors}},
  title = {Agent Loop},
  year = {2026},
  url = {https://docs.openclaw.ai/agent-loop},
  urldate = {2026-07-21}
}

@misc{hermes2026skills,
  author = {{Nous Research}},
  title = {Working with Skills in Hermes Agent},
  year = {2026},
  url = {https://github.com/NousResearch/hermes-agent/blob/main/website/docs/guides/work-with-skills.md},
  urldate = {2026-07-21}
}

@misc{arpahls2026skillware,
  author = {{ARPA Hellenic Logical Systems}},
  title = {Skillware: A Framework for Modular, Self-Contained AI Skills},
  year = {2026},
  howpublished = {GitHub repository},
  url = {https://github.com/ARPAHLS/skillware},
  urldate = {2026-07-21}
}

@article{yang2026federatedskill,
  author = {Yang, Jingbo and Yao, Guanyu and Zhang, Yang and Kompella, Ramana Rao and Liu, Gaowen and Chang, Shiyu},
  title = {{FederatedSkill}: Federated Learning for Agentic Skill Evolution},
  journal = {arXiv preprint arXiv:2606.03143},
  year = {2026},
  eprint = {2606.03143},
  archivePrefix = {arXiv},
  doi = {10.48550/arXiv.2606.03143},
  url = {https://arxiv.org/abs/2606.03143}
}

@article{xu2026skillfab,
  author = {Xu, Anjie and Cai, Yifeng and Li, Yi and Wang, Zixing and Zhang, Zhiyu and Chen, Jingfan and Xu, Ruohan and Wang, Leye},
  title = {{SkillFab}: An Agent-Native Skill Production Platform},
  journal = {arXiv preprint arXiv:2607.03780},
  year = {2026},
  eprint = {2607.03780},
  archivePrefix = {arXiv},
  doi = {10.48550/arXiv.2607.03780},
  url = {https://arxiv.org/abs/2607.03780}
}

@misc{skillsnewapps2026,
  author = {Chen, Le and Wang, Zichang and Zheng, Wenxin and Feng, Erhu and Du, Dong and Xia, Yubin and Chen, Haibo},
  title = {Skills Are the New Apps---Now It's Time for Skill {OS}},
  year = {2026},
  month = feb,
  publisher = {MDPI AG},
  doi = {10.20944/preprints202602.1096.v1},
  url = {https://doi.org/10.20944/preprints202602.1096.v1},
  note = {Preprints.org, version 1, posted February 13, 2026}
}

\end{document}